# Advanced LIGO Two-Stage Twelve-Axis Vibration Isolation and Positioning Platform

## Part 1: Design and Production Overview

**F. Matichard[1,2,*], B. Lantz[3], K. Mason[1], R. Mittleman[1], B. Abbott[2], S. Abbott[2], E. Allwine[5], S. Barnum[1], J. Birch[4], S. Biscans[1], D. Clark[3], D. Coyne[2], D. DeBra[3], R. DeRosa[6], S. Foley[1], P. Fritschel[1], J.A. Giaime[4,6], C. Gray[5], G. Grabeel[5], J. Hanson[4], M. Hillard[1], J. Kissel[5], C. Kucharczyk[3], A. Le Roux[4], V. Lhuillier[5], M. Macinnis[2], B. O'Reilly[4], D. Ottaway[1], H. Paris[5], M. Puma[4], H. Radkins[5], C. Ramet[4], M. Robinson[5], L. Ruet[1], P. Sareen[1], D. Shoemaker[1], A. Stein[1], J. Thomas[4], M. Vargas[4], J. Warner[5].**

**[1] MIT, Cambridge, MA, USA**

**[2] Caltech, Pasadena, CA, USA**

**[3] Stanford University, Stanford, CA, USA**

**[4] LIGO Livingston Observatory, Livingston, LA, USA**

**[5] LIGO Hanford Observatory, Hanford, WA, USA**

**[6] Louisiana State University, Baton Rouge, LA, USA**

* **Corresponding Author:** fabrice@ligo.mit.edu

LIGO Project MIT

MIT NW22-295

185 Albany Street

Cambridge, MA 02139 USA

Phone: +001-617-253-6410

Fax: +001-617-253-7014

**Abstract**

New generations of gravity wave detectors require unprecedented levels of vibration isolation. This paper presents the final design of the vibration isolation and positioning platform used in Advanced LIGO to support the interferometer's core optics. This five-ton two-and-half-meter wide system operates in ultra-high vacuum. It features two stages of isolation mounted in series. The stages are imbricated to reduce the overall height. Each stage provides isolation in all directions of translation and rotation. The system is instrumented with a unique combination of low noise relative and inertial sensors. The active control provides isolation from 0.1 Hz to 30 Hz. It brings the platform motion down to $10^{-11}\ m/\sqrt{Hz}$ at 1 Hz. Active and passive isolation combine to bring the platform motion below $10^{-12}\ m/\sqrt{Hz}$ at 10 Hz. The passive isolation lowers the motion below $10^{-13}\ m/\sqrt{Hz}$ at 100 Hz. The paper describes how the platform has been engineered not only to meet the isolation requirements, but also to permit the construction, testing, and commissioning process of the fifteen units needed for Advanced LIGO observatories.

## 1 Introduction

Physics experiments and precision systems often require a large amount of vibration isolation. Isolators have been developed for a variety of application such as scanning tunneling microscopy [1]-[2], gravitometers [3]-[4], atom interferometric measurements [5], atomic force microscopy [6], colliders [7], and space pointing [8]. Ground based gravity waves detectors have set very stringent requirements in terms of vibration isolation [9]. These instruments use km long interferometers in order to detect strains of space-time caused by astrophysical events [10]. To make gravity waves detection possible, the detector's components must be isolated from all environmental disturbances, including ground motion, which is the dominant disturbance at low frequency.

Seismic isolation concepts and prototypes were developed in early experiments carried out for gravity waves detectors [11]. During the past two decades, several observatories have been built around the world [12]-[16]. Multiple pendulum suspensions equipped with springs blades providing vertical isolation were developed to isolate the optics of the 600 meter long GEO detector located in Germany. Silica fibers are used between the two bottom stages to increase the quality factor and therefore reduce the thermal noise [17]-[18]. This technique in now being used in other gravity waves detectors [19]. The 3 km long VIRGO detector combines 7 meter high inverted pendulums, multi-stage suspended pendulums and inertial control to provide the suitable isolation to all degrees of freedom [20]-[21]. This system called the super-attenuator features natural frequencies as low as 40 mHz. It provides 15 orders of magnitude of isolation at 10 Hz [22]. The 3 km long KAGRA detector in Japan is currently being built underground. The seismic motion in this environment is roughly two orders of magnitude quieter than at the surface level. A combination of seismic attenuation systems and multiple pendulums are used to achieve the isolation requirements [23]. Estimated noises show that the seismic noise will be well under other noise sources in the detection band.

The LIGO observatory based in the US consists of 4 km long detectors, located in Washington State and Louisiana State [12]. After a decade of operation, the initial LIGO interferometers are currently being retrofitted with a new generation of instruments called Advanced LIGO [24]. To provide very high vibration isolation at all frequencies, Advanced LIGO combines Hydraulic Exo-vacuum Pre-Isolators (HEPI), Intra-vacuum Seismic Isolators (ISI) and multistage passive suspensions (SUS) [25]-[27]. The HEPI system is based on the quiet hydraulic actuators and control techniques developed at Stanford [28]-[29]. This active platform provides long-range alignment capability to all directions of translation and rotation. It is used to reject very low frequency disturbances such as tidal and micro-seismic motion [30]. It provides active inertial isolation from 100 mHz to 10 Hz [31]. The ISI platforms feature large optical tables on which the LIGO optics are mounted [32]. Instrumented with low noise instruments, they provide alignment capability and inertial isolation from about 100 mHz to 30 Hz. They also provide passive isolation above a few Hertz to several hundreds of Hertz. Two different types of platform have been developed: the HAM-ISI for the auxiliary optics [33]-[34], and the BSC-ISI for the core optics which require further seismic isolation [35]. The LIGO suspensions holding the interferometer optics are mounted on the ISI systems. These multiple pendulums provide multistage passive isolation. They are equipped with relative sensors and actuators to damp the suspension resonances and to position the interferometer optics. Different types of suspensions have been designed for the different optics used in the interferometers [27].

This paper presents the ISI active platform designed to support the Advanced LIGO core optics suspensions. Installed in the large LIGO vacuum chambers called Basic Symmetric Chambers (BSC), this isolator is often referred to as the BSC-ISI system. It features a two meter wide optical table capable of supporting more than 1000 kg of equipment (optical payload). The design requirement is to provide more than three orders of magnitude of isolation at low frequency, to bring the optical motion down to $10^{-11}\, m/\sqrt{Hz}$ at 1 Hz, and $10^{-12}\, m/\sqrt{Hz}$ at 10 Hz [36]. Five BSC-ISI units per interferometer are necessary to support the core optics (one for the beam splitter, one for each of the two input test masses, and one for each of the two output test masses).

The concept used for the BSC-ISI is based on early prototypes built at JILA in the nineties [37]-[39]. These experiments demonstrated the performance achievable with multi-dof active platforms instrumented with inertial instruments and driven with voice coil actuators through feedback controls. The results obtained motivated the construction of a rapid prototype as a concept for the Advanced LIGO project [40]-[41]. This two-stage system was instrumented with commercial seismometers and voice coil actuators. All six degrees of freedom of each stage were servo controlled. The conclusive results led to the construction of a full-scale two-stage Technical Demonstrator (Tech-Demo) built and tested at the LIGO Stanford facilities [42]. This system had the size and payload capacity required for the Advanced LIGO project. Like the rapid prototype, the Tech-Demo was made of two stages in series, each having six degrees of freedom. Vertical springs were used for the vertical isolation, and flexure rods for the horizontal isolation. An optimal combination of low noise commercial instruments was used for the active control. A unique feature of the Tech-Demo is that it uses “reasonably” stiff springs. Unlike most isolators used in similar applications, the low frequency isolation is entirely achieved actively. All the rigid body natural frequencies are near or above 1 Hz. This system demonstrated that the use of stiffer springs greatly simplifies the assembly, leveling, and commissioning steps, but does not affect the active control performance achievable at low frequency.

After this successful experiment, the “stiff spring” two-stage concept became the baseline design of the In-vacuum Seismic Isolation system needed to support the Advanced LIGO core optics. In 2004, detailed design requirements were defined for the system [43]. A prototype was designed during the following year [44]-[45]. The BSC-ISI prototype was built and tested at MIT between 2006 and 2008 [35]. The lessons learned during the prototyping phase were used to engineer the final design in 2009 and 2010 [46]-[47]. The first unit was successfully tested at the LIGO MIT test facility in 2011. In the past two years, 13 of the 15 units have been built for the LIGO observatories (The last two units are currently under construction).

This manuscript is the first part of a series of two companion papers presenting the BSC-ISI platform for Advanced LIGO. It presents the system and focuses on design and production considerations. The next section gives an overview of the two-stage isolator concept. The third section describes the design of the sub-systems. It highlights the technical challenges related to the design of high precision isolators and details how they have been solved. The fourth section describes design choice trade-offs driven by practical aspects of the production and testing process. The paper concludes with a comparison of the theoretical and experimental transfer functions. The second part of this series of two companion papers focuses on the testing investigation during the prototyping and production phase of this project [48].

## 2 Concept Overview

The final design of the BSC-ISI system is based on the architecture of the system used during the prototyping phase [35]. A conceptual representation is shown in Fig. 1. It represents the structure and the spring components in a schematic section view. Actuators and sensors are not displayed. The system is made of three main sub-assemblies: a base called “Stage 0” and two suspended stages called “Stage 1” and “Stage 2”. The stages are imbricated to minimize the volume occupied. The bottom plate of Stage 2 is the optical table on which the Advanced LIGO equipment is mounted.

Stage 1 is suspended from Stage 0, and Stage 2 is suspended from Stage 1. The spring assemblies provide horizontal and vertical flexibility. They are symbolically represented by helicoids. Their actual shape and location is presented in section 3.5. The springs decouple the stages from each other in all directions of translation (called Longitudinal, Transverse and Vertical in Fig. 1) and all directions of rotation (called Pitch, Roll and Yaw in Fig. 1). The system is designed to minimize the cross couplings between degrees of freedom. In each direction, the system behaves as a two-mass-spring system as illustrated in Fig. 2 for three of the six directions. In Fig. 2 (a), $x_0$, $x_1$ and $x_2$ are the longitudinal motions of Stage 0, Stage 1 and Stage 2. $k_{x1}$, $c_{x1}$ and $m_1$ are the stiffness, damping and mass of Stage 1. $k_{x2}$, $c_{x2}$ and $m_2$ are the stiffness, damping and mass of Stage 2. $f_{x01}$ is the actuation force between Stage 0 and Stage 1. $f_{x12}$ is the actuation force between Stage 1 and Stage 2. Similar notations and subscripts are used in Fig. 2 (b) for the vertical motion and in Fig. 2 (c) for the pitch motion, where the letter $z$ is used to denote the vertical direction, $\theta$ is used for the angular pitch motion, $\tau$ for the torques and $I$ for the quadratic moment of inertia.

In each direction, the system provides passive isolation as described in the system of equations (1)-(3), using the longitudinal direction as an example. In these equations, $x_0$ is the input motion, $x_1$ is the first stage motion and $x_2$ is the second stage motion. The mass, damping and stiffness matrices are called $M$, $C$ and $K$ respectively. The matrices $A$ and $B$ give the influence of the input disturbance and the control forces on the system respectively.

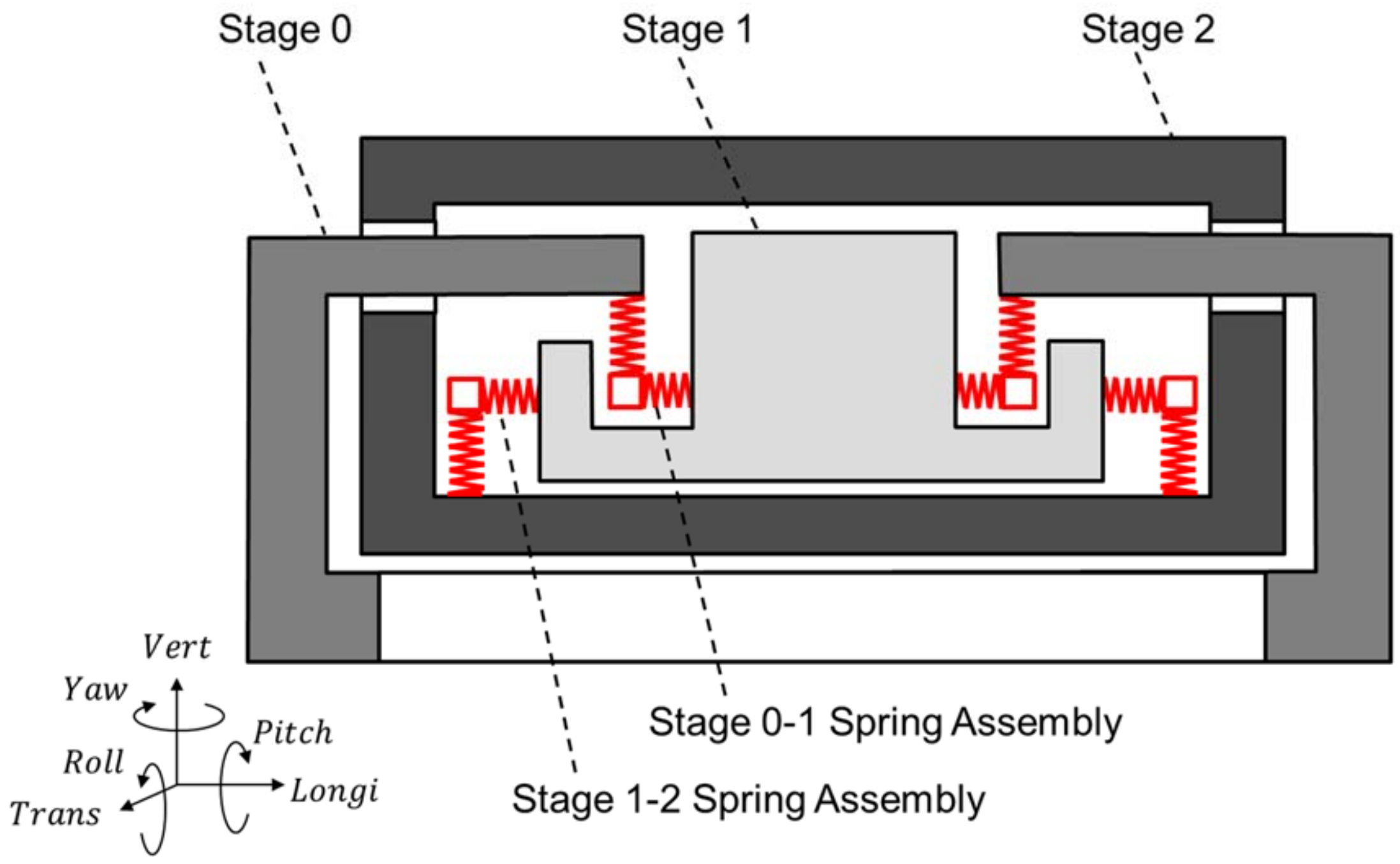

**Fig. 1 Conceptual representation of the BSC-ISI passive components. Sensors and actuators not represented. Springs assembly are symbolically represented by helicoids.**

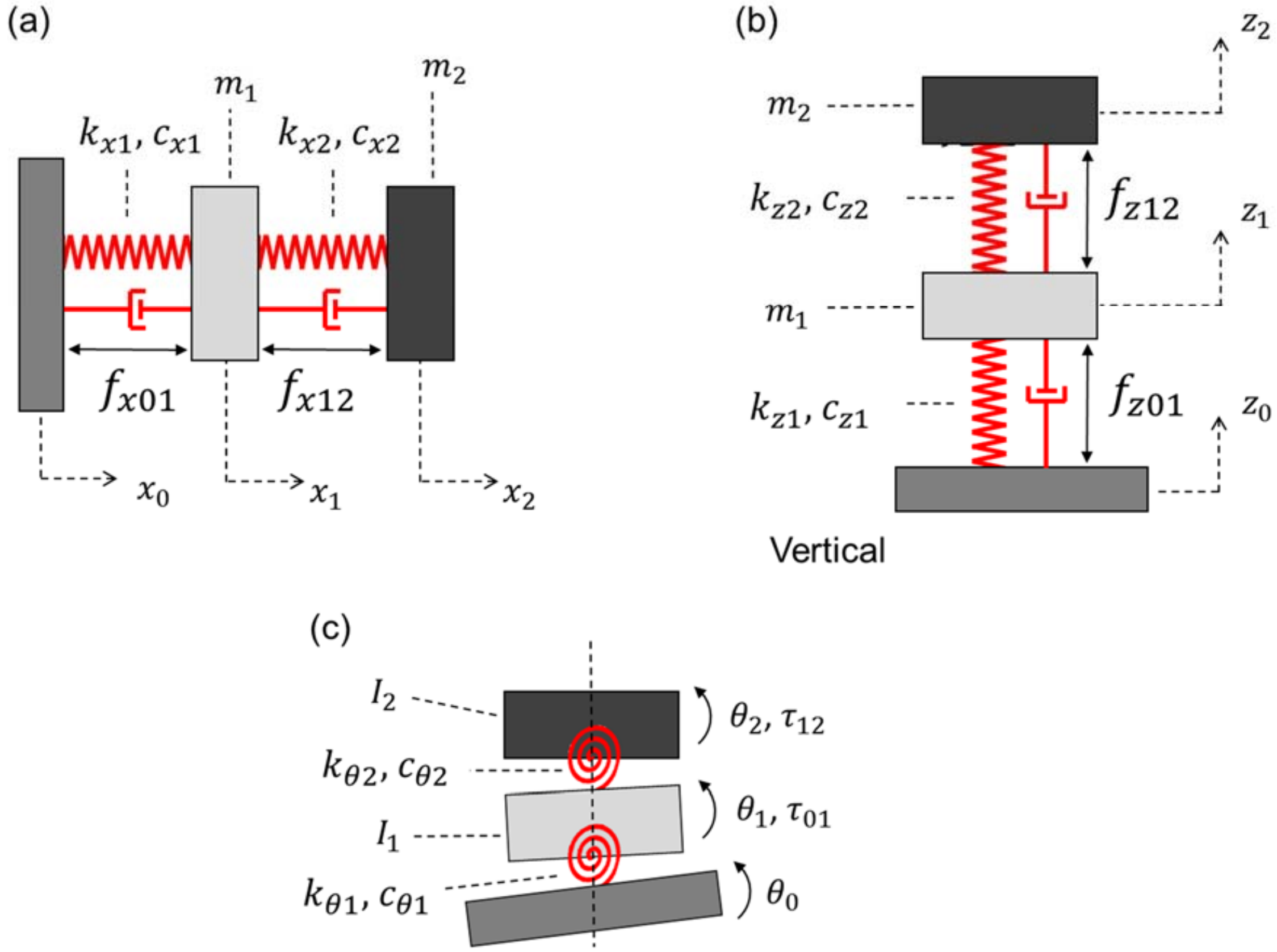

**Fig. 2 Conceptual representation of the passive model along a single axis, for (a) the longitudinal direction, (b) the vertical direction, and (c) the pitch direction.**

$$M\begin{Bmatrix}\ddot{x}_1\\ \ddot{x}_2\end{Bmatrix}+C\begin{Bmatrix}\dot{x}_1\\ \dot{x}_2\end{Bmatrix}+K\begin{Bmatrix}x_1\\ x_2\end{Bmatrix}=A\begin{Bmatrix}x_0\\ \dot{x}_0\end{Bmatrix}-B\begin{Bmatrix}f_{x01}\\ f_{x12}\end{Bmatrix} \tag{1}$$

$$M=\begin{bmatrix}m_1 & 0\\ 0 & m_2\end{bmatrix}, C=\begin{bmatrix}c_1+c_2 & -c_2\\ -c_2 & c_2\end{bmatrix}, K=\begin{bmatrix}k_{x1}+kx_2 & -k_{x2}\\ -k_{x2} & k_{x2}\end{bmatrix} \tag{2}$$

$$A=\begin{bmatrix}k_1 & c_1\\ 0 & 0\end{bmatrix},\ B=\begin{bmatrix}1 & -1\\ 0 & 1\end{bmatrix} \tag{3}$$

At low frequencies, Stage 1 and Stage 2 are controlled actively. Each of the twelve degrees of freedom are controlled independently. For instance, the longitudinal motion of the first mass ($x_1$) is controlled with the force $f_{x01}$ applied between Stage 0 and Stage 1, and the longitudinal motion of the second mass ($x_2$) is controlled with the force $f_{x12}$ applied between Stage 1 and Stage 2. Fig. 3 shows the feedback control block diagram for a single degree of freedom, where $X_\mathrm{n}$ is the degree of freedom under control. It is valid for both stages. The subscript value $\mathrm{n}$ can be $1$ for stage 1, or $2$ for stage 2. Capital letters are used to denote the analysis is performed in the harmonic domain. The stage motion $X_\mathrm{n}$ is disturbed by the input motion $X_{\mathrm{n}-1}$ through the seismic path called $P_s$. It is controlled with the force $F_x$ (Complex amplitude of $f_{x01}$ if analyzing Stage 1, or $f_{x12}$ if analyzing Stage 2) through the force path called $P_F$ .The system’s harmonic responses $P_s$ and $P_F$ can be calculated from theoretical models such as in equations (1)-(3) for the preliminary design phases. Experimental transfer functions are used for the final design of the control loops [48].

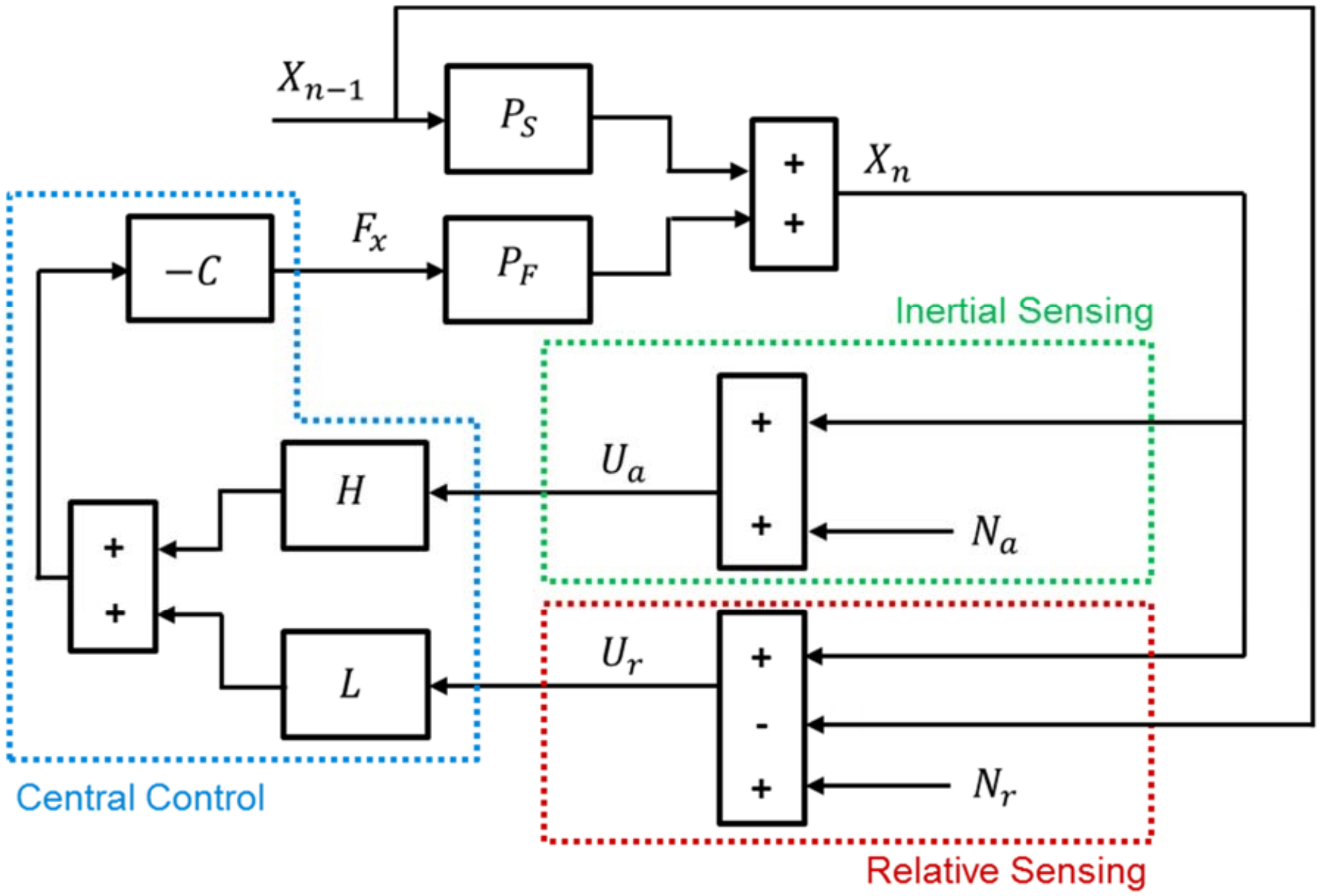

**Fig. 3 Feedback control principle for one degree of freedom.**

On each of the two active stages (Stage1 and Stage 2), two types of sensors are used: inertial and relative sensors. The inertial sensors (geophones or seismometers) are used to measure the stage's absolute motion ($X_{\mathrm{n}}$) which is necessary to provide active seismic isolation (isolation/decoupling from the previous stage). The absolute measurement (inertial instrument signal) is called $U_a$ in Fig. 3. It includes a noise component called $N_a$. This noise is dominant at low frequency where the inertial sensor loses sensitivity. The relative sensors are capacitive gauges measuring the differential motion between the stages ($X_{\mathrm{n}} - X_{\mathrm{n-1}}$). They are used for DC and low-frequency positioning of the stages. The relative measurement is called $U_r$. It includes a noise component called $N_r$. This noise is typically much lower than the inertial sensor noise at low frequencies. Details related to the type of sensors used on each stage are given in section 3. A sensor fusion is used to combine the absolute and the relative measurement. The inertial sensor signal is filtered with the high-pass called $H$. The relative sensor is filtered with the low-pass $L$. The outputs of these filters are summed to form the error signal. A compensator $C$ is used to command the control force $F_x$.

The components of this block diagram are summarized in Eq. (4) to (7), assuming that the absolute measurement ( $U_a$) and relative measurement ( $U_r$) are calibrated in displacements units. This is done practically using digital filters inverting the instruments frequency response. Equation (4) gives the stage motion as a function of the input motion (disturbance) and the control force. Equation (5) gives the control force as a function of inertial measurement ( $U_a$) and the relative measurement ( $U_r$).

Equations (6) and (7) introduce the sensor noise. To simplify the control design, the low-pass and high-pass filters are designed to be complementary [30], as shown in Eq (8). Equation (9) gives the closed loop motion (power spectral density) ${X_\mathrm{n}}^2$ assuming the input motion and the sensor noises are uncorrelated. The first term shows the contribution of the input stage motion ($X_{\mathrm{n}-1}$). The second term shows the contribution of the absolute (inertial) motion sensor noise ($N_\mathrm{a}$). The third term shows the contribution of the relative motion sensor noise ($N_\mathrm{r}$). Assuming large loop gain in the control bandwidth, the amplitude spectral density can be written as shown in Eq. (10). The input motion contribution is filtered by the low-pass filter L. Therefore, the lower the cutoff frequency the better for the isolation, but the high pass-filter $H$ and low-pass filter $L$ must also be adequately designed to minimize the sensor noise contribution. The optimization consists of designing complementary filters that provide both isolation and enough filtering of the instrument’s noise. More details can be found in [30].

$$X_n = P_s X_{n-1} + P_F F_x \tag{4}$$

$$F_x = -C\,(H\,U_a + L\,U_r) \tag{5}$$

$$U_a = X_n + N_a \tag{6}$$

$$U_r = (X_n - X_{n-1}) + N_r \tag{7}$$

$$L + H = 1 \tag{8}$$

$$X_{\mathrm{n}}{}^{2} = \left(\frac{P_s + L\,C\,P_F}{1 + C\,P_F}\right)^2 X_{\mathrm{n-1}}{}^{2} + \left(\frac{H\,C\,P_F}{1 + C\,P_F}\right)^2 N_{\mathrm{a}}{}^{2} + \left(\frac{L\,C\,P_F}{1 + C\,P_F}\right)^2 N_{\mathrm{r}}{}^{2} \tag{9}$$

$$\lim_{(C\,P_F)\to\infty} X_1 = \sqrt{(L\,X_{n-1})^2 + (H\,N_a)^2 + (L\;N_r)^2} \tag{10}$$

## 3 System and Sub-Systems Design Description

### *3.1 System Overview*

The base of the system (Stage 0), the first suspended stage (Stage 1) and the second suspended stage (Stage 2) are shown in grey shades in the Computer Aided Design (CAD) representation in Fig. 4. Stage 1 is suspended from Stage 0 using three sets of blades and flexures. Stage 2 is suspended from Stage 1 using three sets of blades and flexures similar to those used between Stage 0 and Stage 1. One instance of each type of spring sub-assembly is indicated in Fig. 4. The picture shows how Stage 1 and Stage 2 are imbricated to reduce the system’s volume. The two stages’ mass, inertia properties, and the spring’s stiffness are chosen to obtain suitable rigid-body natural frequencies. More details on that are provided in the following sub-sections. The springs are positioned to minimize the cross couplings as explained in the sub-section related to blades and flexures. A BSC-ISI unit undergoing the test process at the LIGO Hanford observatory is shown in Fig. 5. In this picture, 1100 kg of dummy mass is mounted on the top plate to float the stages during the testing phase. Once the testing is completed, the dummy mass removed, the unit is moved to the detector area where interferometer components are attached to the inverted (down-facing) optical table of Stage 2. Details on the system component shapes and features can be found in the top assembly and sub-assembly drawings [49]. The following sub-sections give a detailed description of the sub-systems.

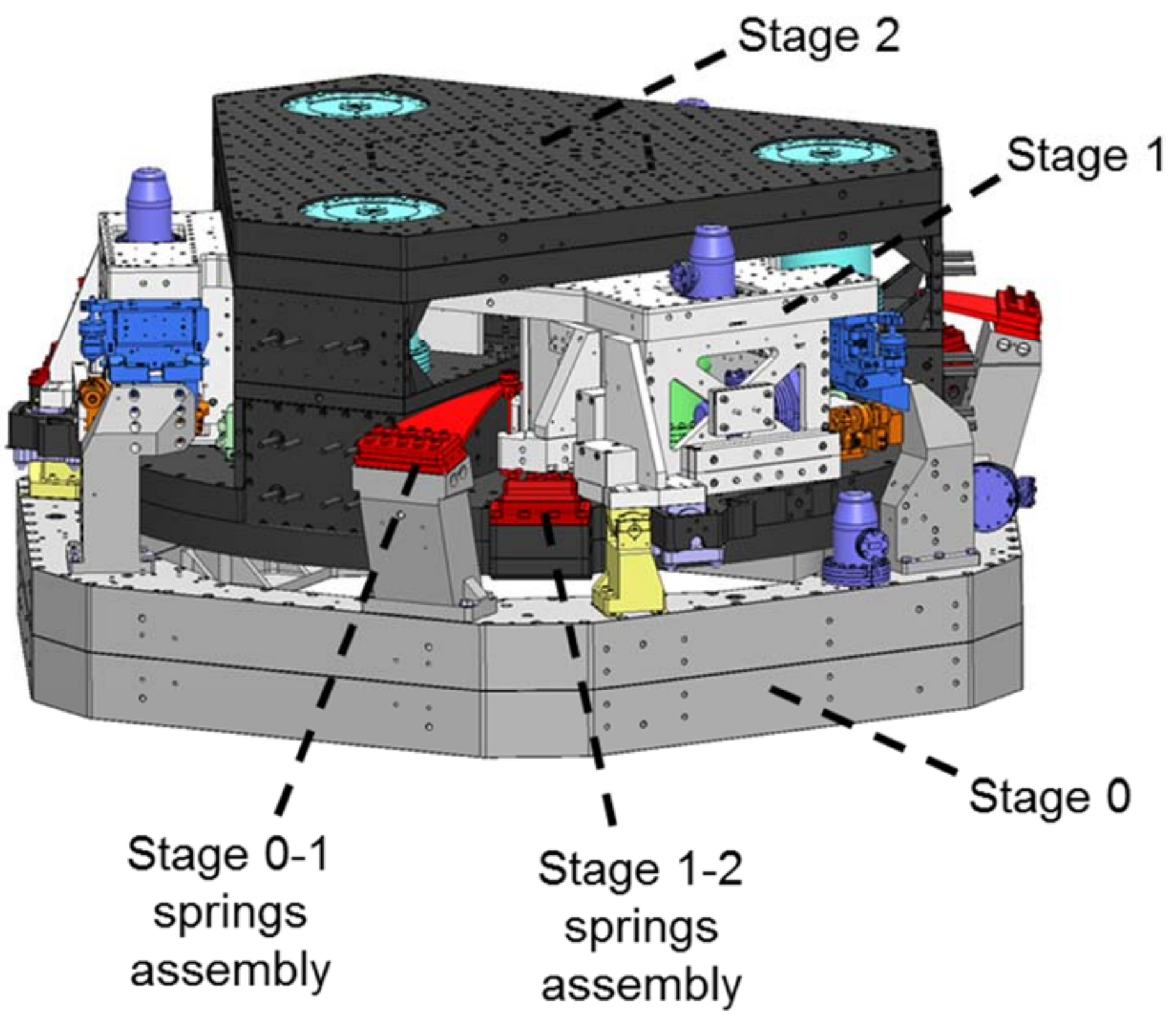

**Fig. 4. CAD representation of the BSC-ISI system.**

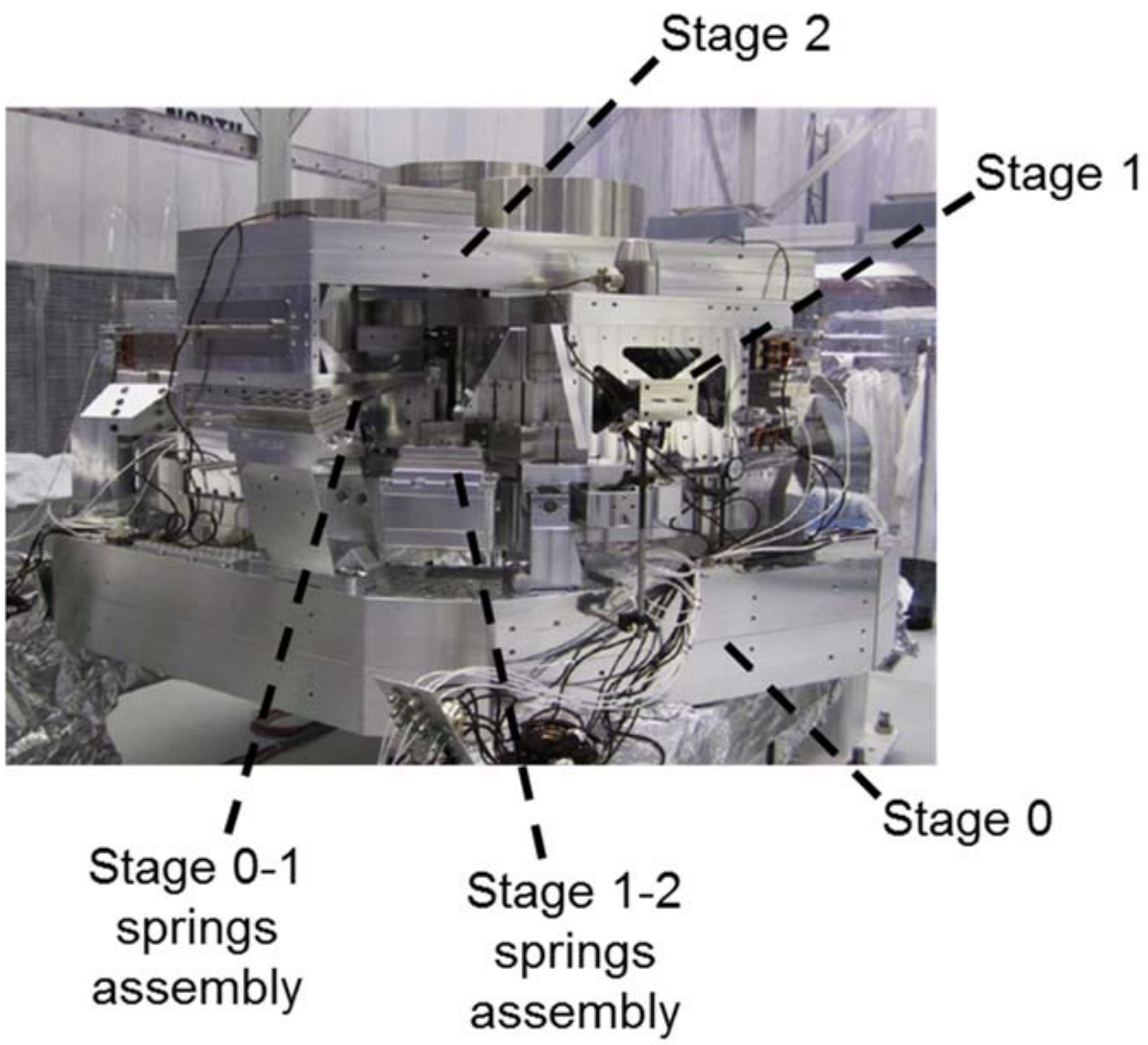

**Fig. 5. A BSC-ISI unit in testing at the LIGO Hanford observatory.**

### *3.2 Stage 0*

Stage 0 is the base of the BSC-ISI system. A CAD representation of Stage 0 is shown in Fig. 6, with the components attached to Stage 0. They are the Stage 0-1 spring assemblies, actuator posts, and motion limiters. The structure of Stage 0 is made of a hexagonal hollow structure approximately 2.4 meters wide and 0.3 meter high. It consists of two monolithic halves bolted together (called top half and bottom half in Fig. 6). Each half structure is machined out bulk aluminum 6061 T6. The inner shapes and webbings have been designed to optimize the structural stiffness, in particular the torsion deformation as illustrated by the blue arrows (angular deformation $\theta(P)$, where $P$ is the static load). The weight of the two half structures combined is 540 kg. The total weight with the components attached to Stage 0 (as shown in Fig. 6) is 885 kg. The 1.6 meter wide hexagonal opening in the inner section of Stage 0 permits access the down-facing optical table of Stage 2.

Machining and assembly errors result in imperfection of the stage locations, spring load, and leveling offsets. The tolerances must, therefore, be adequately chosen and controlled. A machining tolerance of 0.125 mm is specified for the flatness and parallelism of Stage 0 reference surfaces to ensure accurate location and orientation of the sub-assemblies mounted on it. Location pins are used to position the two half-structures relative to each other and to position the components mounted on it. Special care has be given to the press-pin process due to the high friction (the system is ultra-clean to operate in UHV environment, more detail is given in the next sections). Stainless steel hardware is necessary for corrosion resistance during the cleaning process and for UHV compatibility. Preload and torques values in the bolted joints have been calculated to account for the high friction in the clean assemblies. Helicoils made from Nitronic 60 are used in bolted assemblies requiring high preload. Similar machining, tolerancing, positioning, bolting, and cleaning techniques are used for the other sub-assemblies presented in the following sub-sections.

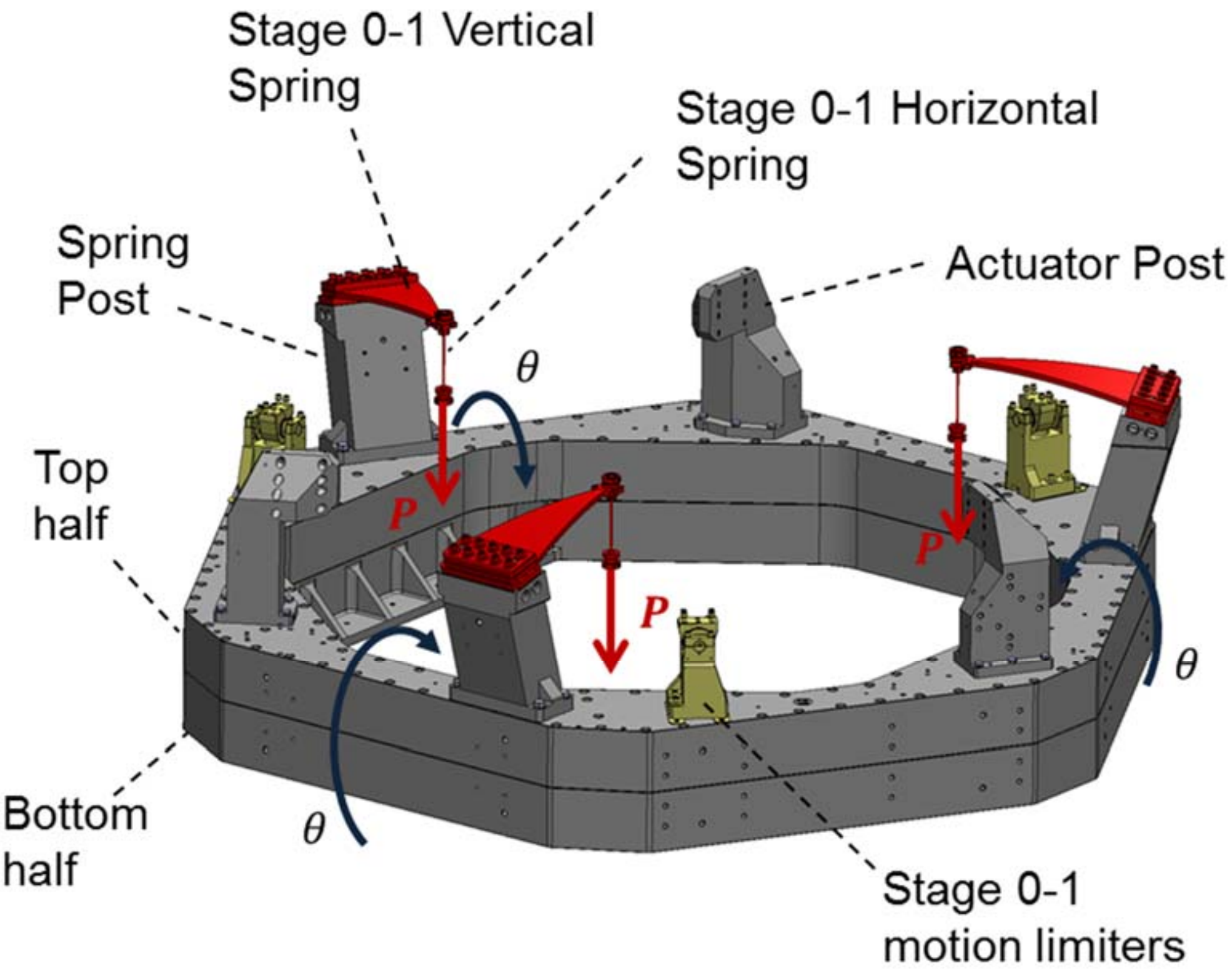

**Fig. 6. A CAD Representation of the Stage 0 structure, Stage 0-1 spring components, actuator posts and motion limiters.**

### *3.3 Stage 1*

A CAD representation of Stage 1 is shown in Fig. 7. It is made of a 1.75 meter-wide three-branch structure. The aluminum structure weighs 550 kg and carries 350 kg of instrumentation (podded sensors and actuators). The structure has been designed to optimize the stiffness over mass ratio. Design and finite element analysis iterations have been carried out to raise the natural frequencies of the structural global bending and torsion modes along the arrows shown in Fig. 7. Related information can be found in [47]. The lowest natural frequency of the metal structure has been measured to be 260 Hz. It drops to 215 Hz when the stage is fully instrumented with all equipment and sub-assemblies. The design is three fold symmetric around the vertical axis. Each of the three branches of the structure is equipped with two relative position sensors, one three-axis seismometer, two geophones, and two actuators.

Two voice coil electromagnetic actuators designed for LIGO by *Planning Systems Incorporated* are mounted in each branch of Stage 1. One actuator is mounted vertically, and one is mounted horizontally. The horizontal actuators are positioned with respect to the spring assembly and the stage center of mass to minimize the cross couplings between horizontal and tilt motion as explained in the section on flexure rods. The coil is mounted on *Stage 0* to improve the heat conduction path out of the vacuum chamber. The magnet assembly is mounted on *Stage 1*. The magnets are positioned in pairs of North-South and South-North dipoles to reduce the magnetic field which escapes the actuator assembly. The actuator has a return yoke to minimize the escaped magnetic field which can interfere with the sensitive equipment surrounding the platform.

These actuators produce 40 N/Amp, with a coil resistance of 6.5 Ohms. Coil drivers capable of supplying a +/- 20 V range are used to drive the actuators. More information on the coil drivers is given in section 3.6. The maximum force delivered by each actuator is approximately 125 N. The first pole induced by the impedance of the coil is at 32 Hz. The dynamic response of the actuator is characterized in [50].

A CAD representation of the actuator assembly is shown in Fig. 8 and a picture of an actual assembly is shown in Fig. 9. Tooling bars (not shown) are used to lock the coil and the magnet components during the assembly process. Aluminum brackets are used to connect each half of the actuator to the stage. Copper bars are used to conduct the heat out of the assembly. Spherical support and washers are used in the joints to reduce over-constrains between the coil and the magnets occurring during the assembly process.

Capacitive gauges supplied by *Microsense* are used to sense the relative motion between Stage 0 and Stage 1. These sensors are collocated with the actuators as shown in Fig. 8. The sensor gauges are attached to Stage 0 so that their wires do not introduce unwanted friction on the isolated stage. The aluminum sensor targets are attached to Stage 1. They are diamond turned to obtain a very smooth finish necessary to minimize the cross-sensitivity to lateral motion. A modulation/demodulation scheme is used to sense the motion. The *Stage 0-1* sensors range of sensing is +/- 1 mm. Their noise floor is $2 \times 10^{-10}\ m/\sqrt{Hz}$ at 1 Hz.

The sensor electronic boards are mounted outside of the vacuum system. Tri-ax cables are used to minimize the cable capacitance and radiation.

One three-axis seismometer is used in each of the three branches of Stage 1. They are 240 seconds period instruments (T240) supplied by *Nanometrics*. One of the two horizontal axes is oriented tangentially, and the other radially. These sensors are very low noise instruments at low frequency. The noise of these inertial sensors is in the range of $2 \times 10^{-9}\ m/\sqrt{Hz}$ at $100\ mHz$.

In order to be used in ultra-high vacuum, each seismometer is mounted in a sealed chamber, as shown in Fig. 10. The chamber is custom-made for this application. It is made of Stainless Steel 304. The welded parts are subjected to a thorough leak-check process to validate the fabrication. For this test, the chamber is filled with Helium, and subjected to a RGA scan. For the final assembly, the pod containing the instrument is filled with Neon, which is used as a tracker for leak detection during operations. Additionally, each pod is instrumented with pressure sensors to help identify a faulty pod in case of a leak. The pod used for the T240 has been designed to minimize the volume between the instruments and the sealed chamber, and therefore reduce noise disturbances related to gas currents induced by temperature gradients.

Stage 1 is also equipped with six passive inertial sensors. They are L4C geophones supplied by *Sercel*. There is one instrument mounted horizontally and one mounted vertically in each branch. The instruments are encapsulated in custom made sealed chambers as previously described for the T240 seismometers. A CAD representation of the L4C sealed assembly is shown in Fig. 11. The instruments are equipped with low-noise custom-made pre-amplifiers and pressure sensors mounted on the instrument inside their sealed chamber. The instrument sensor noise is $2 \times 10^{-11}\ m/\sqrt{Hz}$ at $1\ Hz$ and $10^{-12}\ m/\sqrt{Hz}$ at $10\ Hz$. The T240 seismometers and L4C geophones are combined for very low noise broadband inertial sensing. More information on the sensor fusion is given in [48]. Fig. 12 shows all the components in one branch of stage 1. It also shows the front door and vibration absorbers used to damp the structural modes (they were not displayed in Fig. 7 to better show the instrument location in the branch).

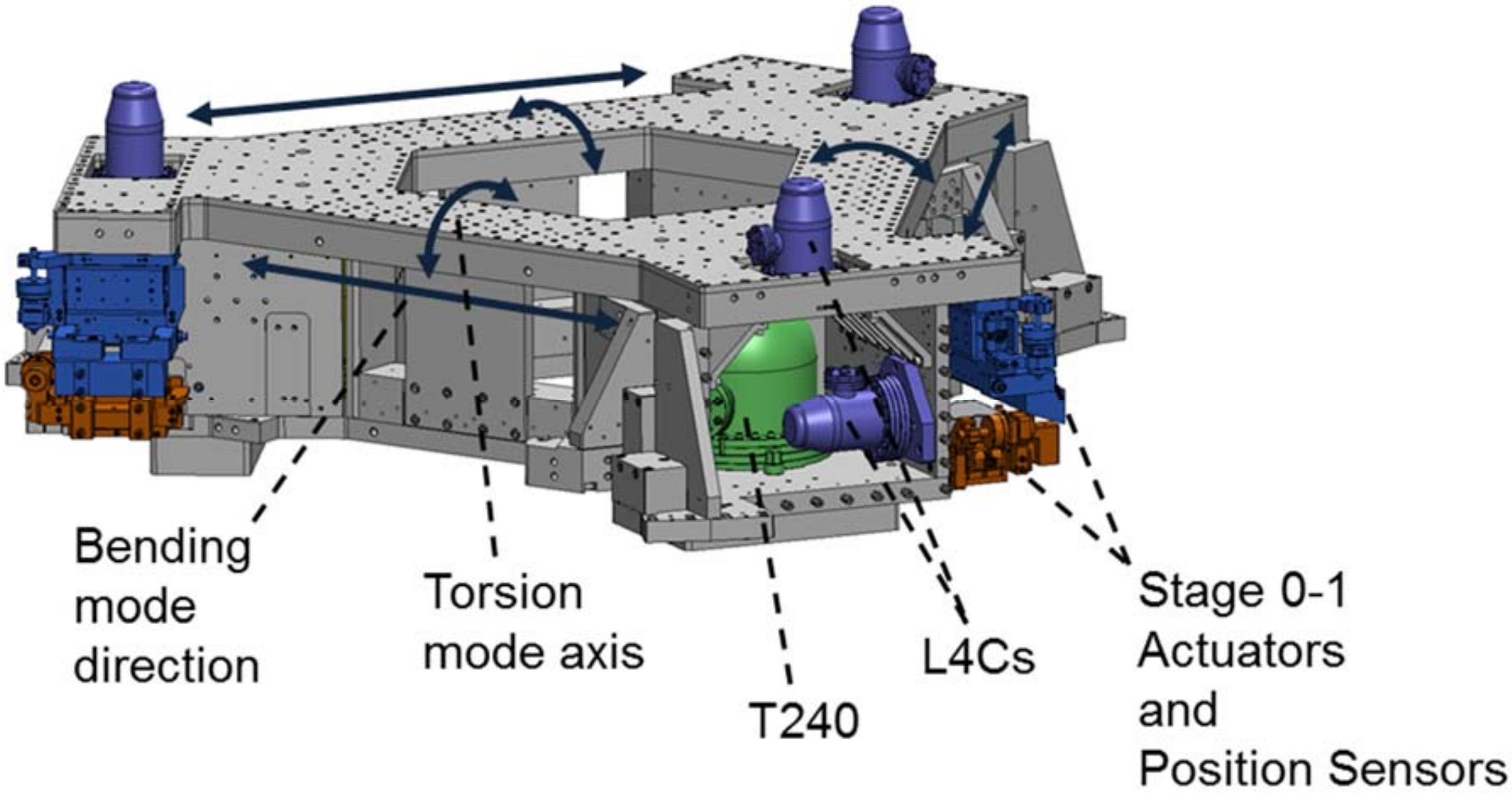

**Fig. 7 A CAD Representation of Stage 1 (Front door not shown).**

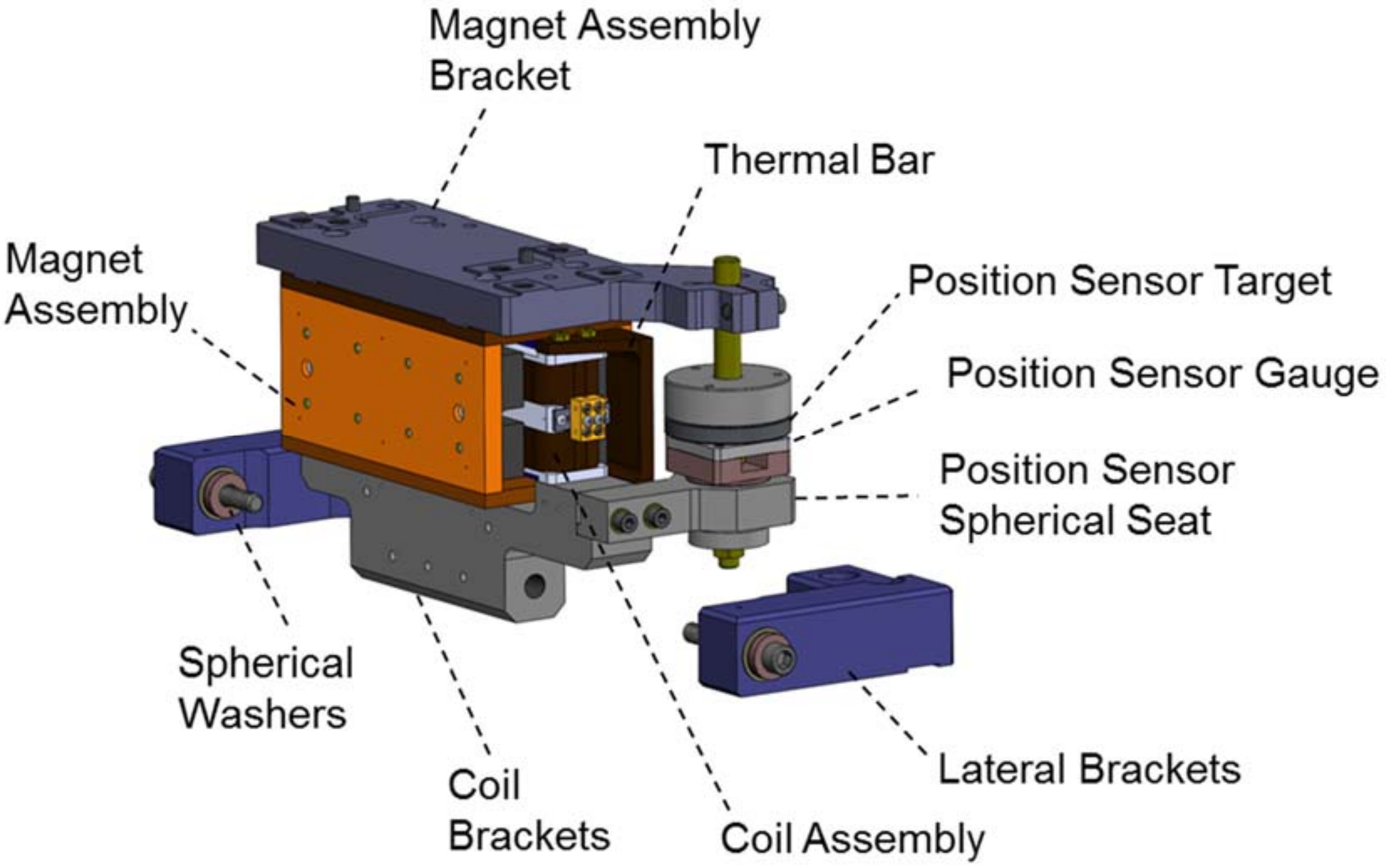

**Fig. 8. A CAD representation of the Stage 0-1 actuators and capacitive position sensors assembly.**

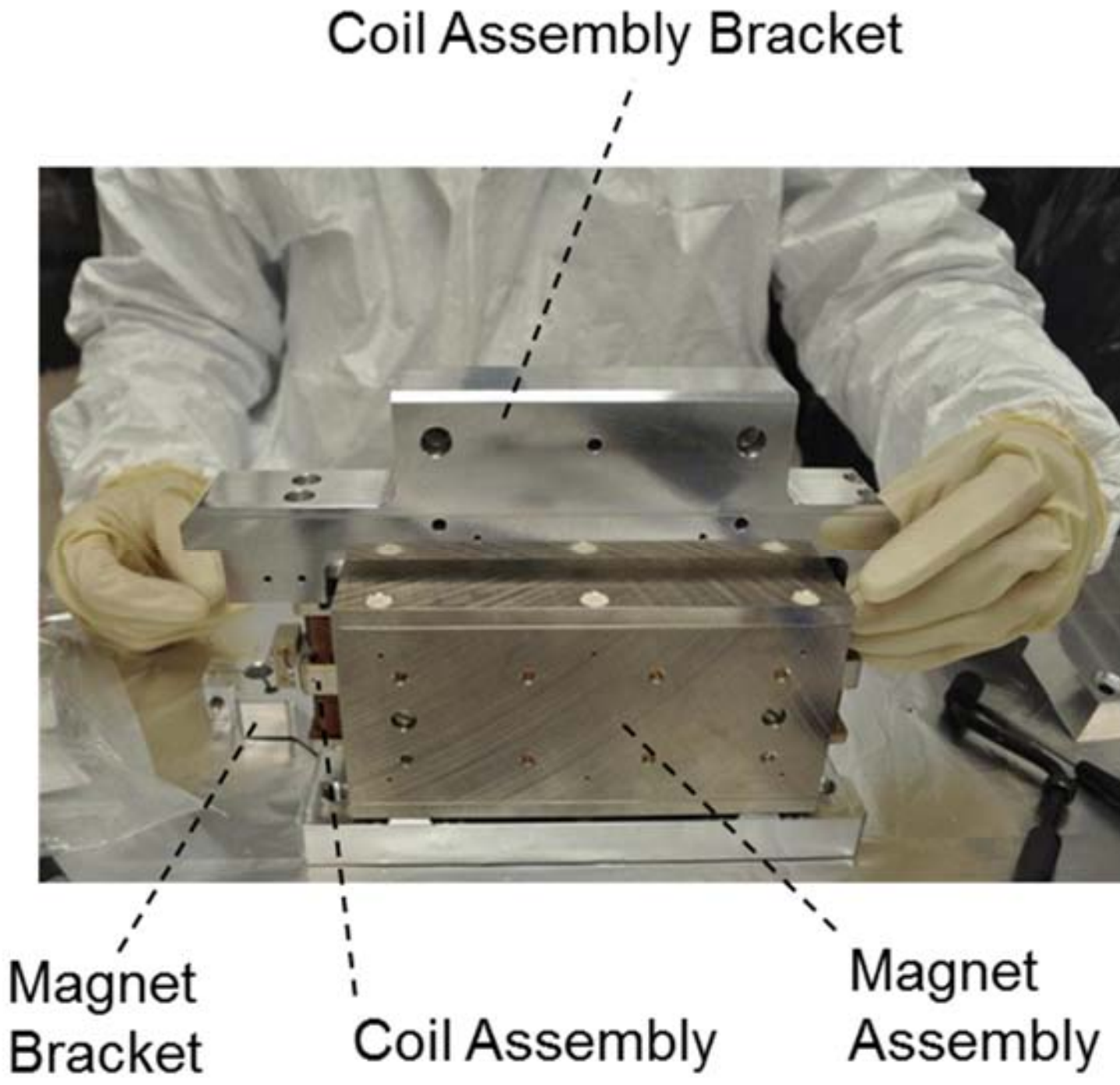

**Fig. 9. Stage 0-1 actuator assembly.**

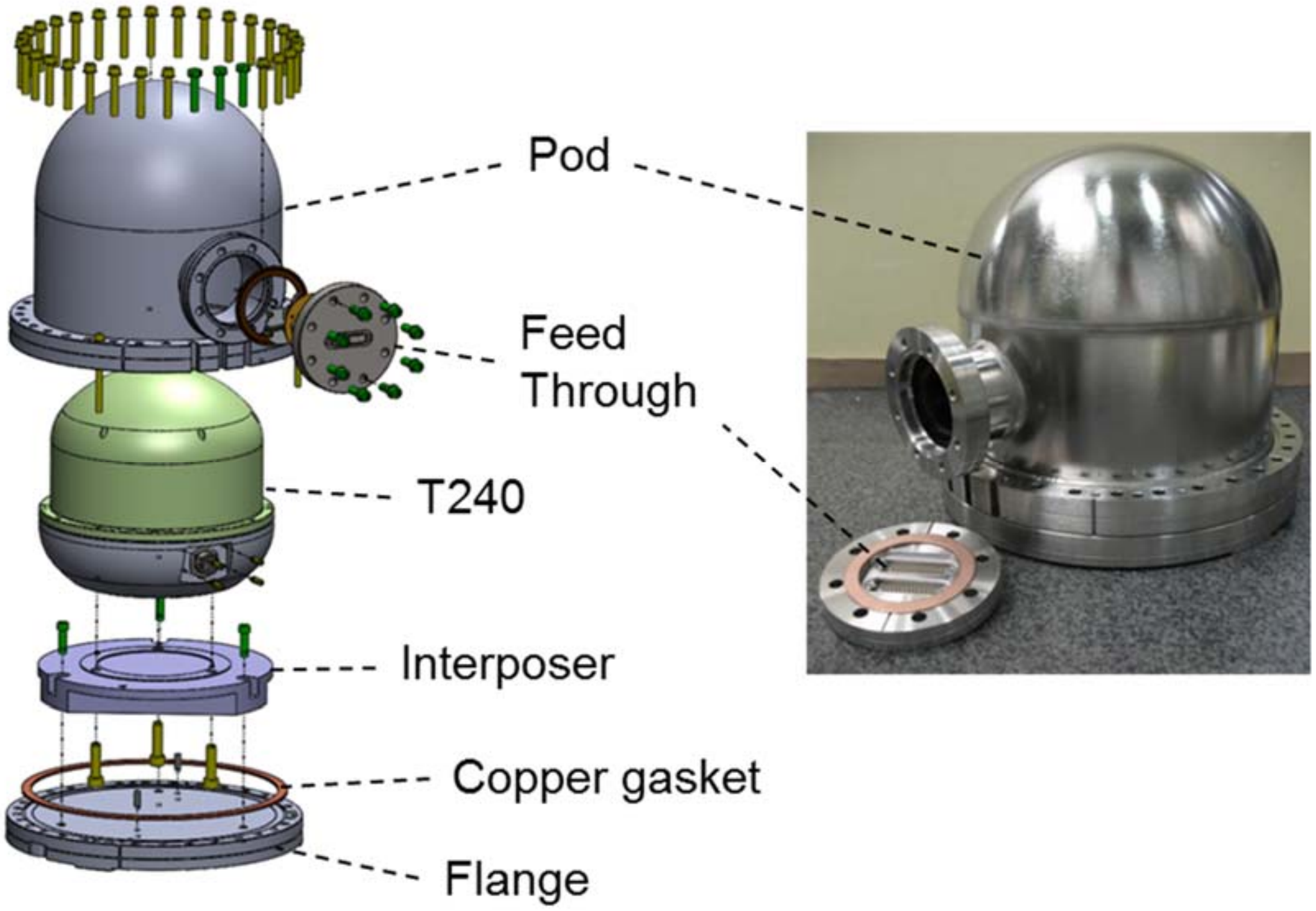

**Fig. 10. Custom made sealed chamber to use the Trillium T240 in ultra-high-vacuum.**

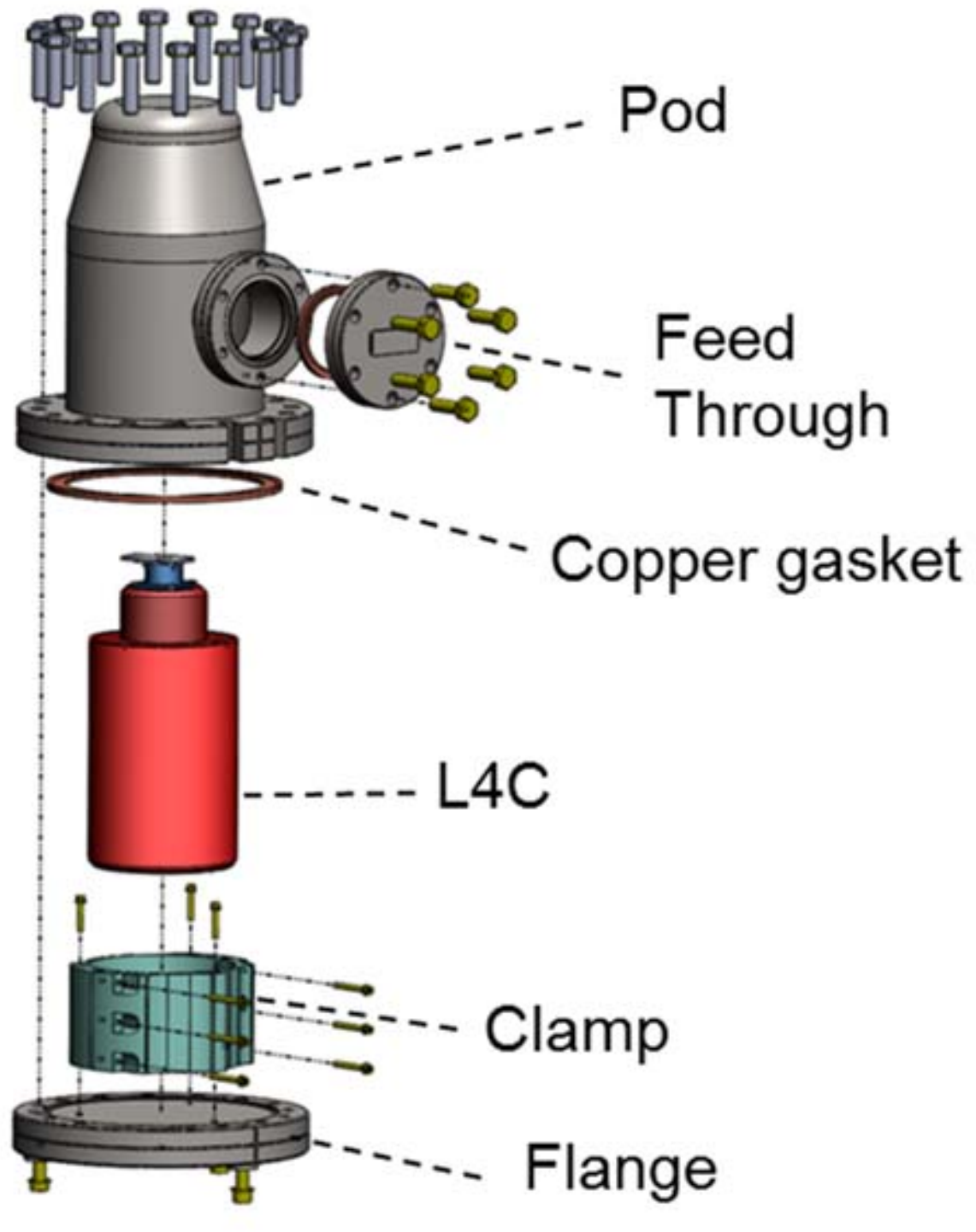

**Fig. 11. Custom made sealed chamber to use the Sercel L4C in ultra-high-vacuum.**

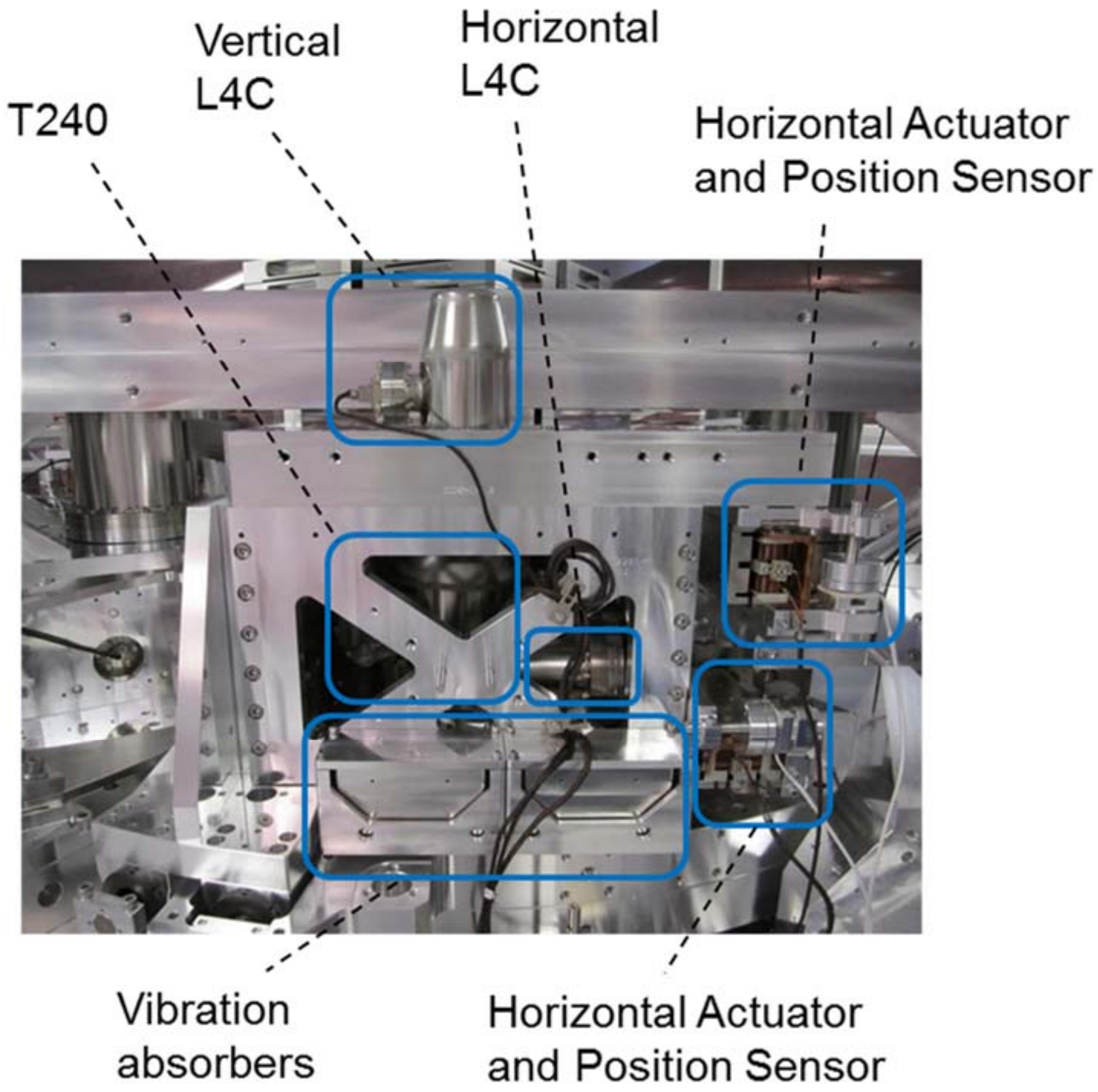

**Fig. 12. Front view of one of the three branches of Stage 1.**

### *3.4 Stage 2*

A CAD picture of Stage 2 is shown in Fig. 13. The 1.85 meter-wide aluminum structure weighs 1350 kg. The total mass of Stage 2 including the podded sensors, actuators, springs, and motion limiters is 1750 kg. Additionally, it supports 1200 kg of equipment, including 350 kg usually configured as ballast. The Stage 2 threefold symmetric structure is built so as to encompass Stage 1. The assembly process of the imbricated stages is detailed in [51]. The bottom plate of Stage 2 features an optical table with a 25 mm x 25 mm hole pattern. Like the two other stages already presented, Stage 2 has been optimized for overall stiffness and local rigidity where the instruments are mounted. It is suspended from Stage 1 using three sets of blades and flexures similar to those used on Stage 1. These components are described in the section 3.5.

In each of its three corners, Stage 2 is instrumented with two actuators, two relative sensors, and two inertial sensors. Fig. 14 and Fig. 15 show the actuators. They are similar to those described for Stage 1. The coils are mounted on Stage 1 to provide a shorter conduction path for the heat to transfer out. It also reduces the risk of a mechanical shortcut through the actuator cabling. Since Stage 1 provides sufficient positioning capability, smaller actuators are used on Stage 2. These actuators develop 27 N/Amp. The coil resistance is 10 Ohms. The maximum force delivered by each actuator is approximately 55 N. The first pole induced by the impedance of the coil is at 50 Hz. The dynamic response of the actuator is characterized in [50].

The Stage 1-2 position sensors are collocated with the Stage 1-2 actuators. The gauges are mounted on Stage 1 and the targets on Stage 2. The Stage 1-2 position sensor construction is identical to those used for Stage 0-1, but they are four times more sensitive, at the cost of four times less sensing range. Their noise floor is $6 \times 10^{-11}\, m/\sqrt{Hz}$ at $1\, Hz$. The low noise performance of these instruments is necessary for the positioning control not to compromise the seismic isolation of Stage 2.

The inertial sensors used for Stage 2 are passive GS13 geophones supplied by *Geotech*. The instrument flexures have been replaced by custom made beryllium-copper notch-style flexures [52]. This modification makes the instrument more robust to shock, and eliminates the need for locking the instrument during transportation and installation. The instruments are equipped with low noise pre-amplifiers carefully chosen to minimize the noise in the control bandwidth [53]. The instrument sensor noise is $8 \times 10^{-12}\, m/\sqrt{Hz}$ at $1\, Hz$ and $4 \times 10^{-13}\, m/\sqrt{Hz}$ at $10\, Hz$. Fig. 16 shows pictures of the GS13's pod assembly. The aluminum external shell of the instrument has been replaced by a mu-metal can to shield the instruments against magnetic fields. Like the inertial instruments of Stage 1, the GS13s are encapsulated in custom-made sealed chambers equipped with pressure sensors, filled with Neon, leak checked after assembly and monitored during operations.

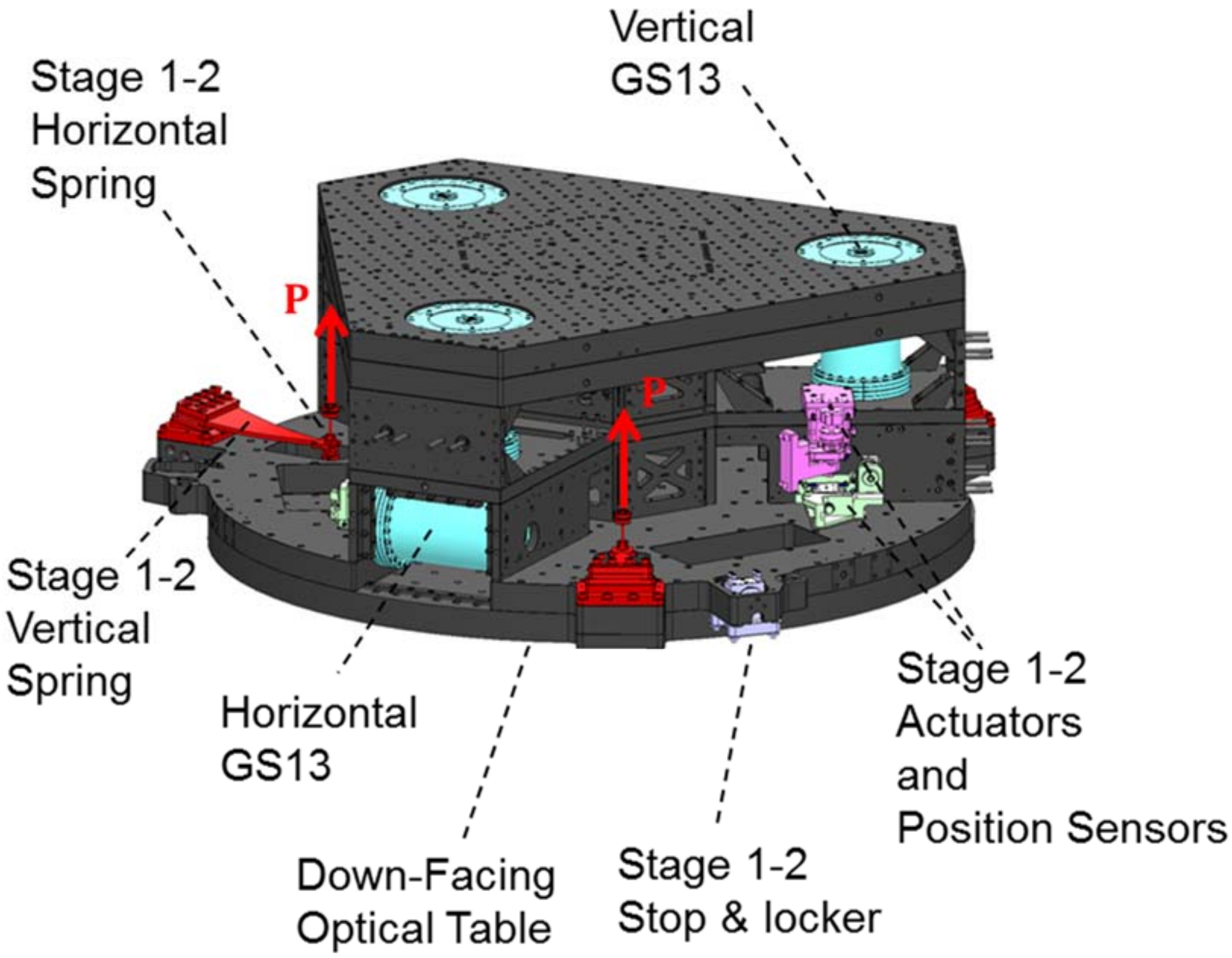

**Fig. 13. A CAD Representation of the Stage 2 Assembly.**

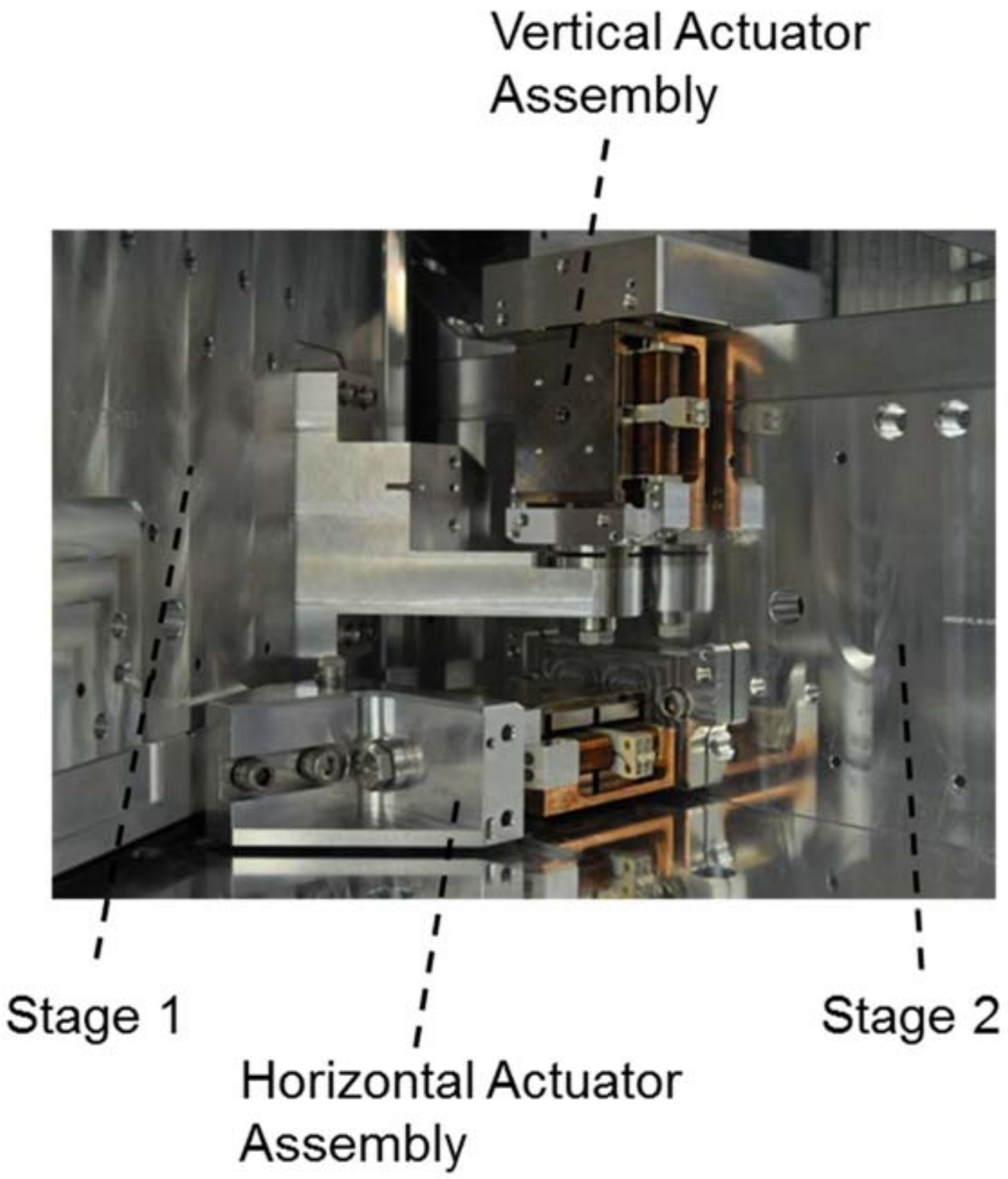

**Fig. 14. Stage 1-2 actuator assembly.**

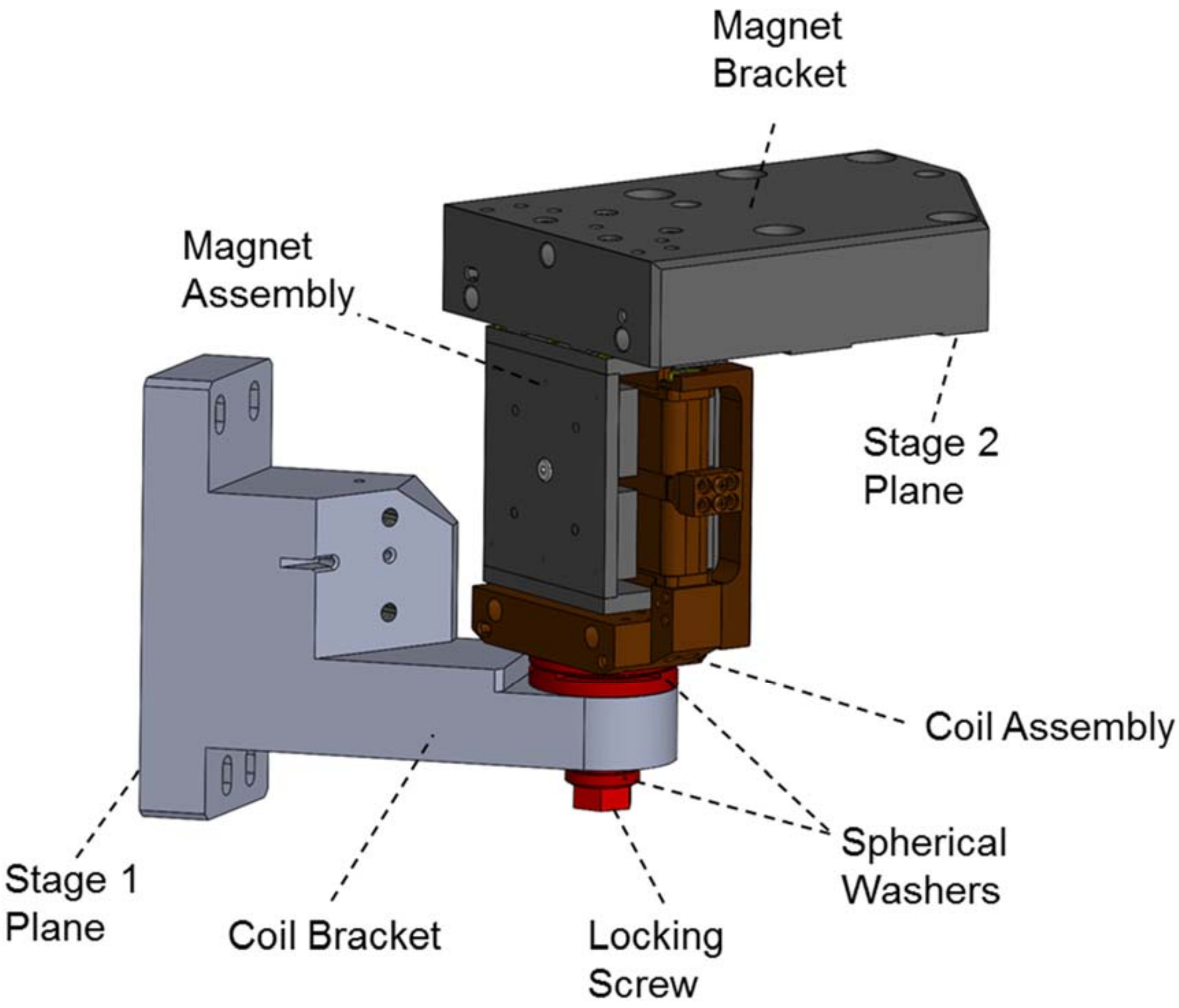

**Fig. 15. Stage 1-2 vertical actuator CAD representation.**

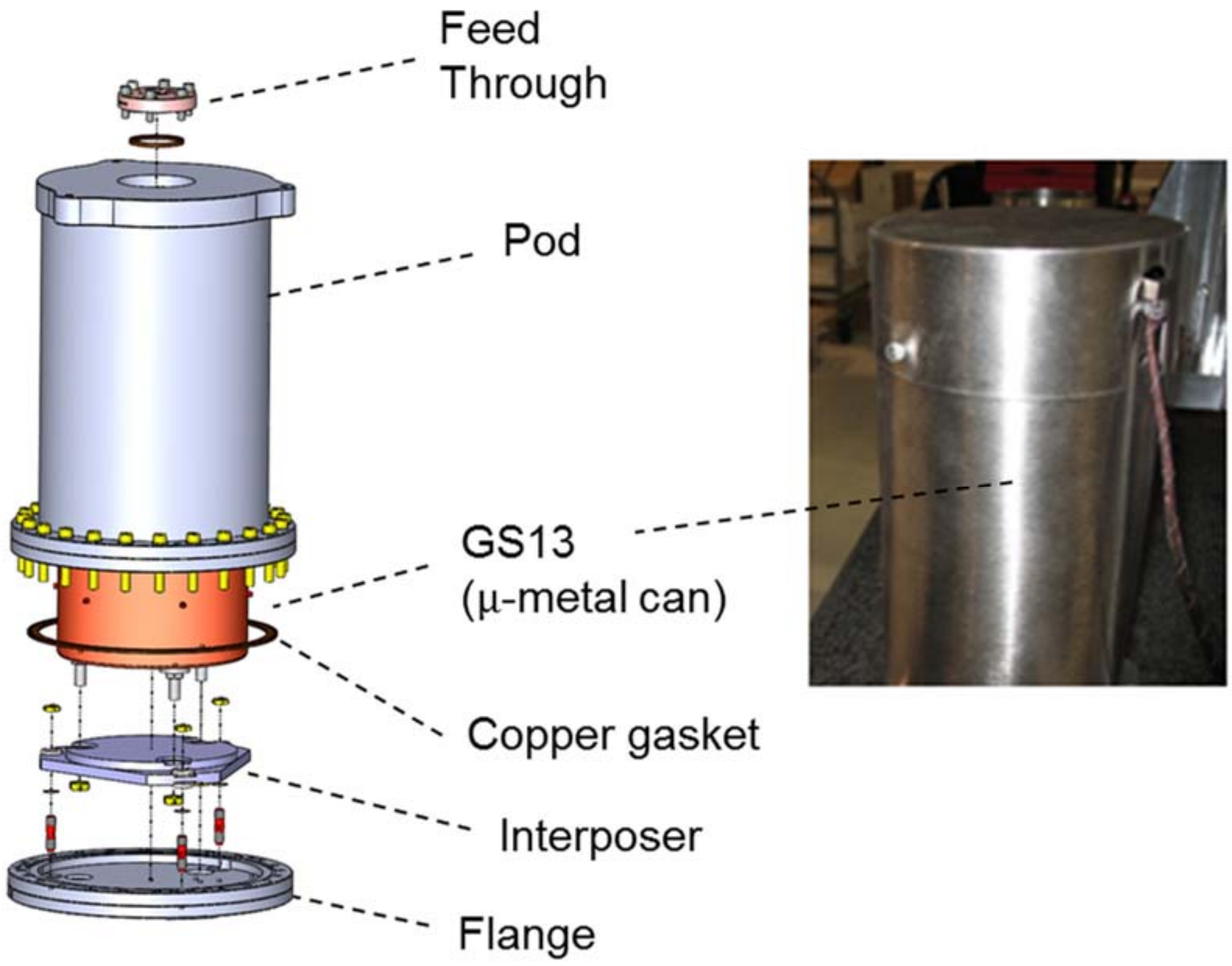

**Fig. 16. GS13 equipped with mu-metal shielding and sealed in a custom pod for use in ultra-high-vacuum.**

## *3.5 Spring Components*

### 3.5.1 Springs assembly overview

The design and engineering of the spring components is very critical to both the performance and operability of a vibration isolation system. The rigid-body natural frequencies must be low enough to provide adequate passive isolation, but the springs must not be too compliant so that the system remains easy to commission and operate. LIGO experience acquired during prototyping of vibration isolation systems showed that rigid-body natural frequencies in the 1 Hz to 7 Hz range provide an excellent compromise.

To reduce the complexity of the control strategy and to facilitate operation, the BSC-ISI system has been engineered to minimize the couplings between the degrees of freedom in the Cartesian basis. As a result, the system behaves as a two-spring-mass system in each direction of translation and rotation as explained in the introduction. The spring components have been designed to obtain in-phase rigid-body modes in the 1 Hz to 2 Hz range (the two masses moving in phase at the resonance), and out-of-phase rigid-body modes in the 5 Hz to 7 Hz range (the two masses moving out of phase at the resonance).

Three blade-flexure assemblies are used on each stage, and positioned symmetrically at 120 degrees around the vertical axis. The blades are designed to provide the vertical flexibility, and the flexure rods are designed to provide the horizontal flexibility. A symbolic cross-section representation of one of the three sets is shown in Fig. 17. It shows one Stage 0-1 spring assembly and one Stage 1-2 spring assembly. The blades are flat in the un-deformed state, and curved with a constant radius of curvature in the loaded configuration as shown in Fig. 17. The stage 0-1 blade center of curvature is located under the blade (with respect to the ascending vertical axis), and the stage 1-2 blade center of curvature is located above the blade to reduce the volume occupied by the springs assembly.

The blades, the flexures, the horizontal actuators and the stage's center of mass are positioned relatively to each over to minimize the cross couplings. Fig. 17 illustrates how the static tilt-horizontal coupling has been minimized. The Stage 0-1 blade's tip and the Stage 0-1 horizontal actuator plane are located at a distance $\lambda_1$ from the flexure's tips. The distance $\lambda_1$ is chosen so that the horizontal actuator force ($F_1$) produces a horizontal motion ($x_1$), but no rotation. The distance $\lambda_1$ is function of the flexure geometric parameters, material, and the vertical load $P$. More details are given in the section on flexures. Similar positioning is done for the stage 1-2 components (location $\lambda_2$, horizontal force $F_2$, horizontal motion $x_2$). A cross section of the Stage 0-1 spring assembly is shown in the CAD representation in Fig. 18, and a cross section of the Stage 1-2 spring assembly is shown in Fig. 19.

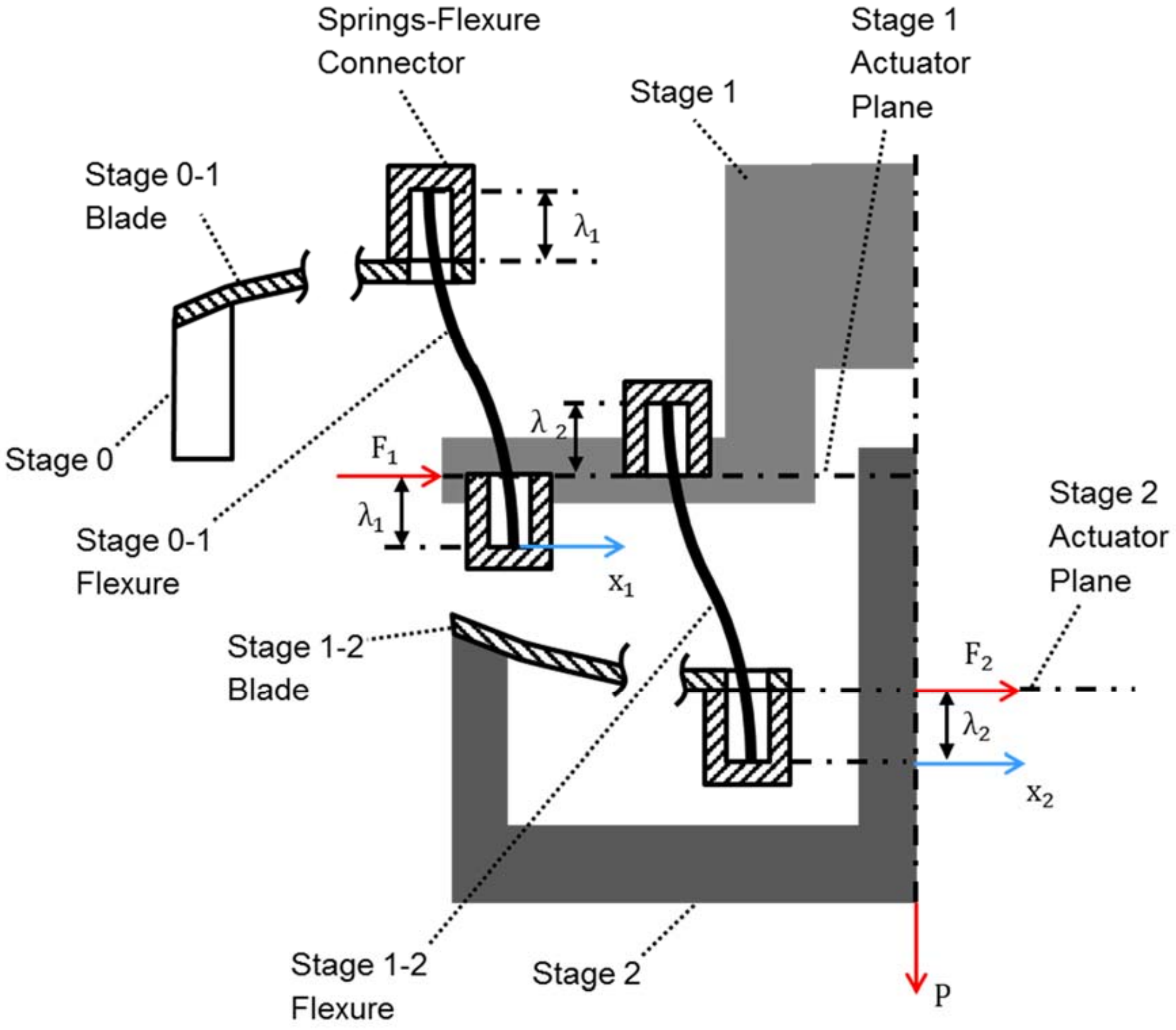

**Fig. 17. Conceptual representation of the spring assembly.**

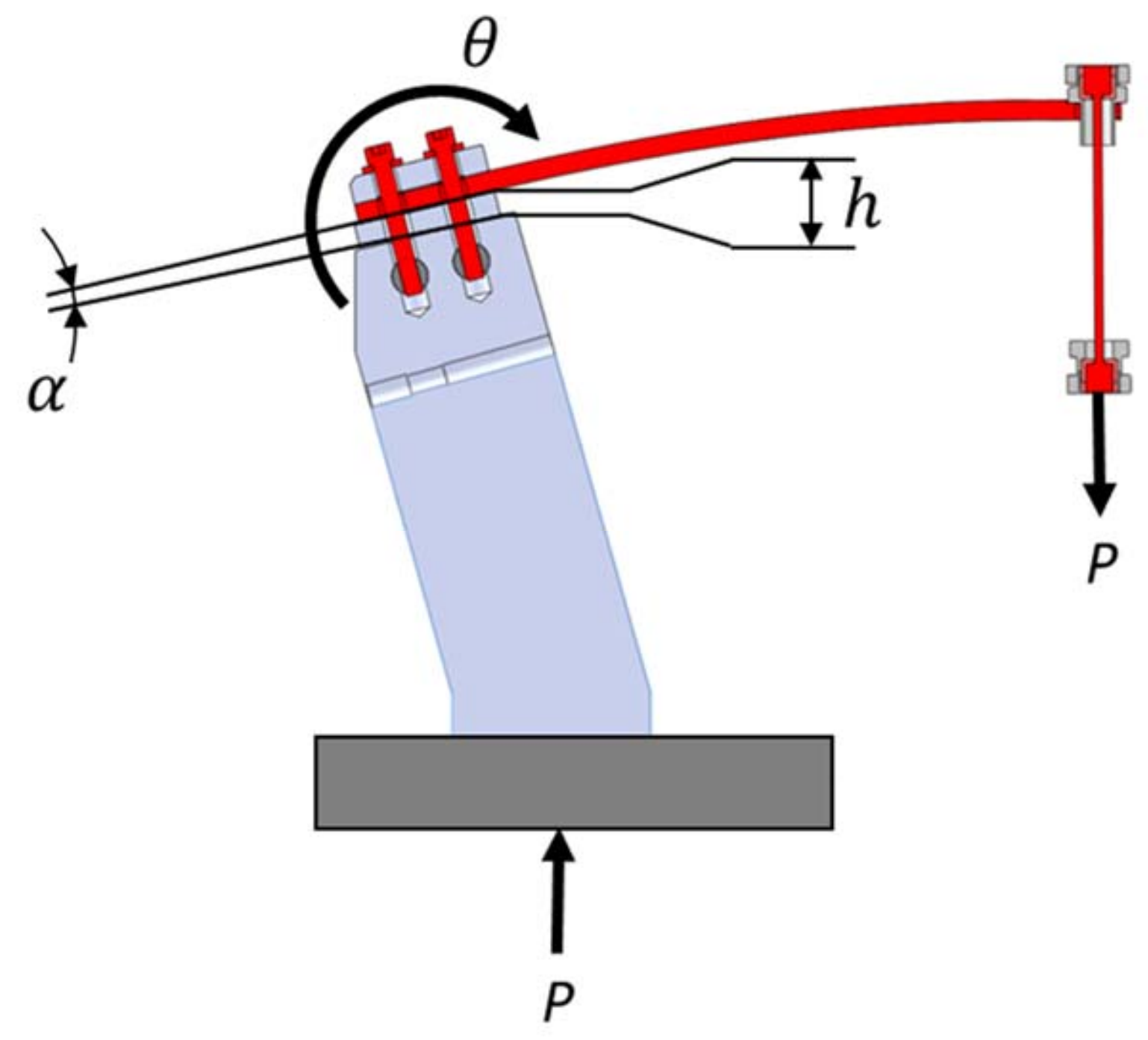

**Fig. 18. Stage 0-1 spring assembly.**

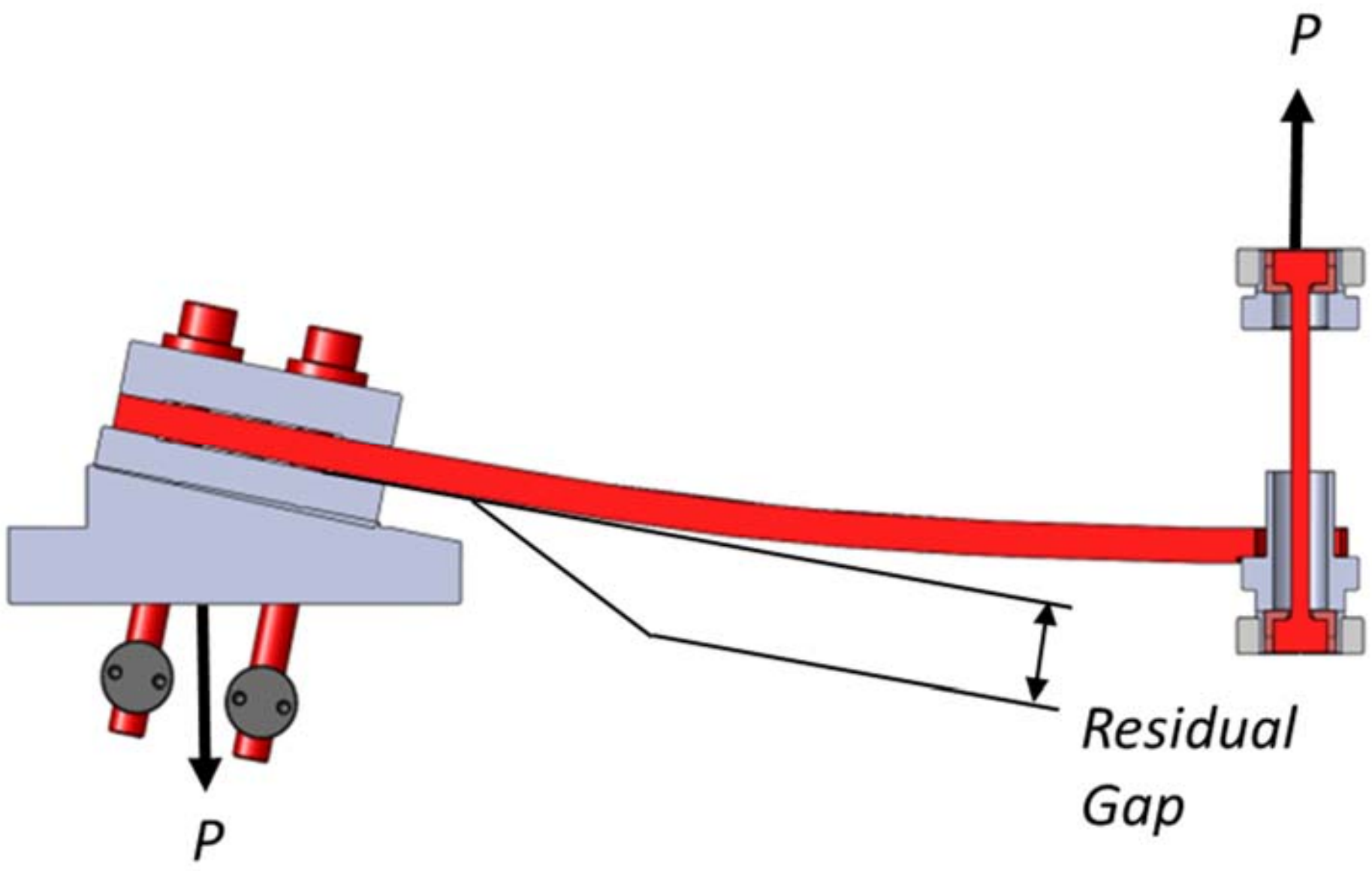

**Fig. 19. Stage 1-2 spring assembly.**

### 3.5.2 Blades providing vertical flexibility

Bernoulli beam theory as given in Eq. (11)-(12) can be used to relate the blade's triangular geometry with the spring constant, where $E$ is the young modulus, $I$ is the quadratic second area moment of inertia, $w$ is the vertical deflection, $x$ is the horizontal location, $P$ is the normal load, $l$ is the length of the blade, $b$ is the width at the root, and $h$ is the thickness. Equation (13) gives the stress $\sigma$ at the surface of the blade. Due to the triangular shape, it is independent of the location $x$. Stress levels are checked through non-linear finite element analysis (FEA) to account for large displacements and to estimate accurately the stress concentration near edges and boundary conditions [54]. Equation (14) give the tip deflection $w$ and Eq. (15) gives the spring's stiffness at the tip of the blade. These parameters are also checked through non-linear FEA. The spring components are made of maraging steel. They are nickel plated to avoid corrosion during the cleaning and assembly process. Tuned mass dampers are mounted on the blades to damp the internal resonances [55]. Table 1 summarizes the blades parameters, analytical calculations and FEA results.

$$EI\ \frac{\partial^2 w}{\partial x^2} = P\ (l - x) \tag{11}$$

$$I = \frac{b\ h^3}{12}\frac{(l - x)}{l} \tag{12}$$

$$\sigma = \frac{6\ P\ l}{b\ h^2} \tag{13}$$

$$w = \frac{6\,P\,l}{E\,b\,h^3}\,x^2 \tag{14}$$

$$k_z = \frac{E\,b\,h^3}{6\,l^3} \tag{15}$$

The clamps are designed to not provide unwanted flexibility (clamp flexibility noted $\theta$ in Fig. 18). Particular attention has been given to minimize the gap between the clamps and the root of the blade that could potentially result in friction and hysteresis effects. Such a residual gap is illustrated in Fig. 19 and shown in Fig. 20 for Stage 1-2 blades. Silver plated screws and Nitronic 60 barrel nuts are used to reduce the friction in the bolted assemblies and increase the preload in the joints. Jigs were constructed to measure the friction in the clean assemblies and correlate the results with FEA analysis.

The geometry of the spacers at the root of the blade must be controlled with accuracy to not affect the system's equilibrium position. The launch angle $\alpha$ (machined to ¼ mrad angular tolerance) and the height $h$ (machined to $25\,\mu m$ tolerance) of the clamps have been used as degrees of freedom during the testing phase to bring the system to its nominal mass and equilibrium position. They are shown in Fig. 18 for stage 0-1 blades. Once the first production units clamp parameters had been adequately tuned, all subsequent units were machined with the same set of parameters.

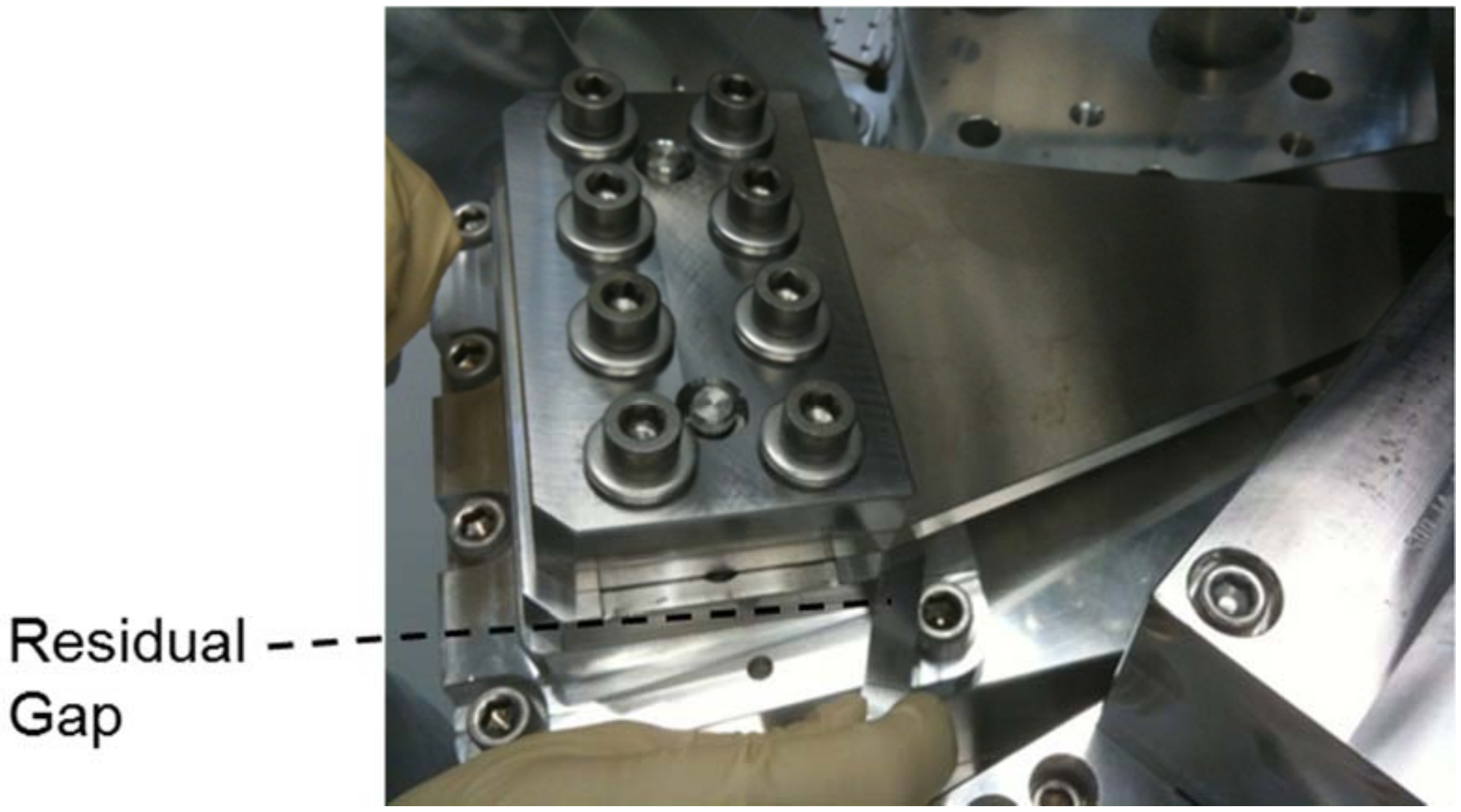

**Fig. 20. Clamp gap inspection: a gap between the blade and its mount can arise if the load force over comes the pre-load force in the bolted assembly. Friction in the ultra-clean bolted assembly and pre-load are carefully controlled.**

**Table 1: Blades Properties**

| Symbol | Name | Stage 0-1 | Stage 1-2 |
|---|---|---|---|
| $h$ | Thickness | $13.87 \times 10^{-3}$ m | $12.14 \times 10^{-3}$ m |
| $b$ | Base width | 0.216 m | 0.162 m |
| $l$ | Length | 0.429 m | 0.322 m |
| $P$ | Load per blade | 12000 N | 9125 N |
| $k_z$ | Stiffness, Analytical | $2.28 \times 10^{5}$ N/m | $2.72 \times 10^{5}$ N/m |
| | Stiffness, FEA | $2.17 \times 10^{5}$ N/m | $2.57 \times 10^{5}$ N/m |
| $w(l)$ | Deflection, Analytical | $52.6 \times 10^{-3}$ m | $33.6 \times 10^{-3}$ m |
| | Deflection FEA | $55.4 \times 10^{-3}$ m | $35.3 \times 10^{-3}$ m |
| $\sigma$ | Stress, Analytical | 750 MPa | 740 MPa |
| | Stress, FEA | 1050 MPa | 1200 MPa |

### 3.5.3 Flexure rods

The flexure rods are attached to the blade's tip as shown in the schematic representation in Fig. 21 (a) for the Stage 0-1 assembly, where $F_1$ is the actuator horizontal force, $P$ is the normal load, and $u_1$ is the tip deflection. Beam equations are used to calculate the distance $\lambda_1$ so that the horizontal force $F_1$ produce no rotation [56]. It can be summarized as follows. Fig. 21 (b) shows the flexure motion parameters and external forces. The flexure's deflection is $u(x)$. The axial load is $P$. The force, translation motion, torque and rotation at the top tip are $F_0$, $u_0$, $\tau_0$ and $\theta_0$ respectively. The force, translation motion, torque and rotation at the bottom tip are $F_1$, $u_1$, $\tau_1$ and $\theta_1$ respectively. Equation (16) gives the beams' cross section equilibrium, where $E$ is the young modulus, $I$ is the quadratic moment of inertia, and $z$ is the vertical location. The general solution is given in Eq. (17) where the constant values $a_0$, $a_1$, $a_2$ and $a_3$ are function of boundary conditions, and $K$ is given in Eq. (18). The solution can be written as a function of the external forces and motions at the tips as shown in the system in Eq. (19), where $k_{tt}$ is the translational stiffness, $k_{rr}$ is the rotational stiffness, $k'_{rr}$ and $k_{tr}$ are cross coupling terms. This system can be solved to find the location of the forces $F_0$ and $F_1$ that produces pure translational motion *(*$\theta_0 = \theta_1 = 0$*).* This specific location, sometimes called "zero-moment-point" location is noted and given in Eq. (20). The horizontal actuator mid-plane is positioned at a distance $\lambda_1$ from the bottom tip where the actuator force $F_1$ produces a pure translation. The blade is positioned at a distance $\lambda_1$ from the top tip where the reaction force from the blade $F_0$ will not rotate the rod. Similar positioning is done for the Stage 1-2 blades. The flexures geometrical parameters are chosen to provide both suitable stiffness and peak stress value. The maximum axial stress can be estimated as shown in (21), where the maximum deflection $u_1$ is constrained by motion limiters. The Stage 0-1 motion limiters allow +/- 0.5 mm of motion. The Stage 1-2 motion limiters allow +/- 0.25 mm of motion. The flexure rods properties are summarized in Table 2.

$$EI\ \frac{\partial^2 u}{\partial z^2} = -\tau_1\ + F_1\ z + P\ (u - u_1) \tag{16}$$

$$u(z) = a_0 + a_1\ z + a_2 \cosh(K\ z) +\ \alpha_3 \sinh(\ K\ z) \tag{17}$$

$$K = \sqrt{\frac{P}{E\ I}} \tag{18}$$

$$\begin{Bmatrix} F_0 \\ \tau_0 \\ F_1 \\ \tau_1 \end{Bmatrix} = \begin{bmatrix} k_{tt} & k_{tr} & -k_{tt} & k_{tr} \\ k_{tr} & k_{rr} & -k_{tr} & k'_{rr} \\ -k_{tt} & -k_{tr} & k_{tt} & -k_{tr} \\ k_{tr} & k'_{rr} & -k_{tr} & k_{rr} \end{bmatrix} \begin{Bmatrix} u_0 \\ \theta_0 \\ u_1 \\ \theta_1 \end{Bmatrix} \tag{19}$$

$$\lambda = \frac{k_{tr}}{k_{tt}} = \frac{1}{K}\ \tanh \frac{KL}{2} \tag{20}$$

$$\sigma = \frac{4\ P}{\pi\ d^2} + \frac{P\ \lambda\ d\ u_1}{2\ \ I\ (L - 2\ \lambda\ )} \tag{21}$$

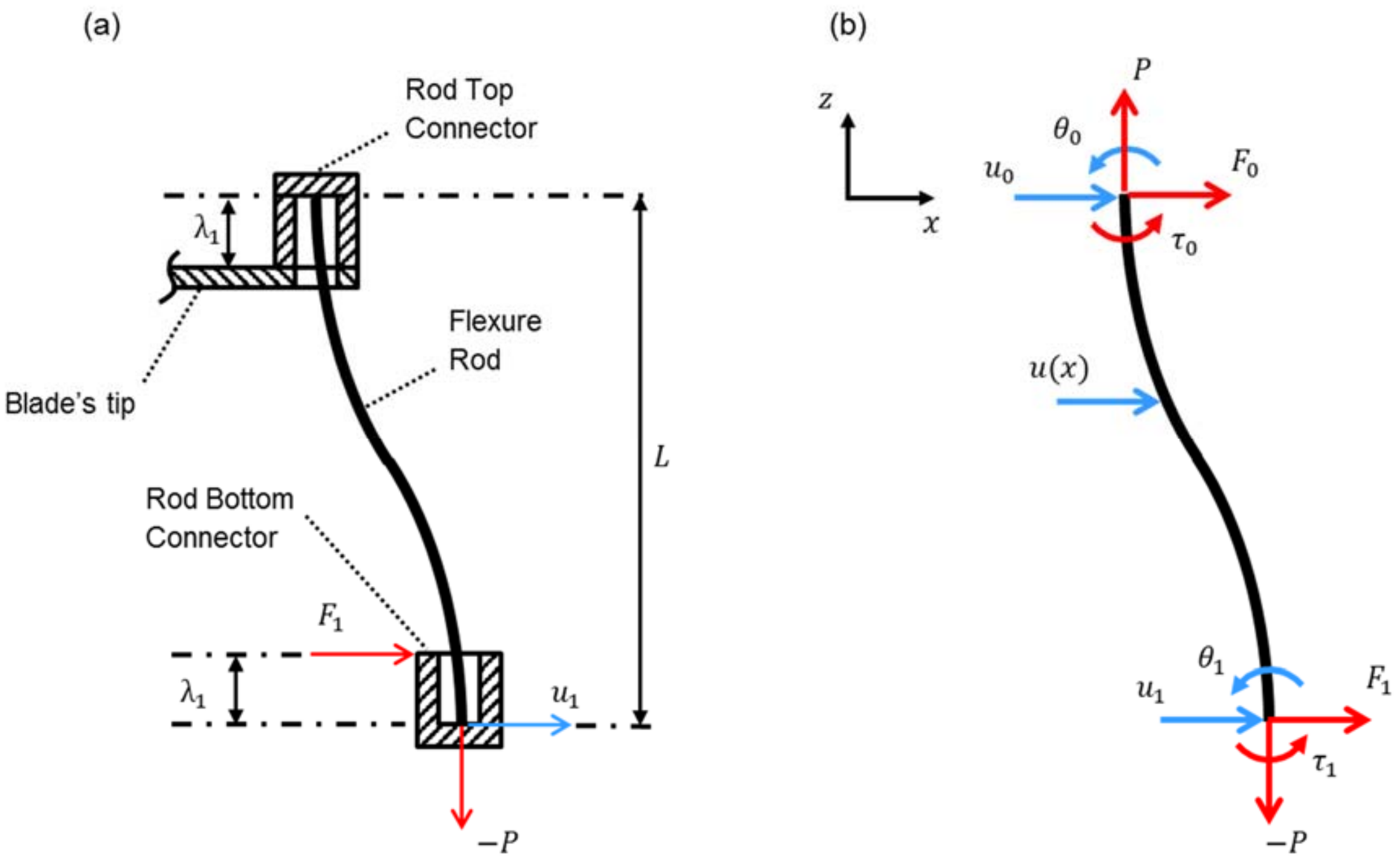

**Fig. 21. (a) Conceptual representation of the flexure rod attached to the blade. (b) Force and position parameters.**

**Table 2: Flexure rods properties**

| Symbol | Name | Stage 0-1 | Stage 1-2 |
|---|---|---|---|
| $d$ | Diameter | 5.92 x $10^{-3}$ m | 6.35 x $10^{-3}$ m |
| $L$ | Length | 0.189 m | 0.117 m |
| $P$ | Load per Flexure | 12000 N | 9125 N |
| $kx$ | Stiffness, Analytical | 9.41 x $10^{4}$ N/m | 2.32 x $10^{5}$ N/m |
| $\lambda$ | Zero Moment Point | 30.6x $10^{-3}$ m | 32.9x $10^{-3}$ m |
| $w(l)$ | Maximum Deflection | 0.6x $10^{-3}$ m | 0.3x $10^{-3}$ m |
| $\sigma$ | Stress, Analytical | 522 MPa | 471 MPa |

### *3.6 Electronics and Computing*

The electronics of high performance active vibration isolation systems must be very carefully designed to allow such systems to perform at design sensitivity. For instance, the pre-amps of passive geophones will likely limit the self-noise of the instrument-preamp pair [57]-[58]. The amplification and filtering stages must be sufficiently low noise to not compromise the measurement. The dynamic range must be carefully considered. For that, adequate gain and filtering must be chosen to maintain the signal above the quantization noise without saturating the analog to digital converter (ADC). A detailed knowledge of the input motion and environmental conditions is therefore necessary for the design of the amplification stages. Switchable gains and whitening stages are often necessary to adjust the settings as a function of a change of control state or fluctuation of input conditions.

Fig. 22 shows a simplified signal flow diagram for the BSC-ISI system. It indicates the sensitivity of the instruments mounted on the platform. The L4C and GS13 geophones are equipped with custom-made pre-amplifying boards. Detailed information regarding the selection of the op-amp can be found in [53]. The boards are mounted directly on the sensor connector, inside the sealed pod. These electronic boards also host the pressure sensors used to detect a possible leak.

Low-noise interface chassis are used to filter, collect and send the signals to the ADC. The parameters of amplification and filtering are specific to each type of sensor (though they are called “*Ampli*” in Fig. 22 ). The analog gains and whitening filters can be switched by the control system in order to adjust the dynamic range to match operating conditions. The signals are digitized at 64 kHz using 16 bit +/- 20V ADC cards. A third order Chebyshev with a notch near the sampling frequency is used for the anti-aliasing filter. The signals are decimated to the digital controller sampling frequency at 4 kHz. The digital control is based on the LIGO CDS real time code [59]. An Epics database [60] is used for communication between the front-end real-time controller and the control room machines. The operator interface is built in MEDM code [61]. The controller output is up-sampled to the DAC cards frequency at 64 kHz. Low-noise voltage amplifiers (*coil drivers*) are used to drive the magnetic actuators. Actuator voltages and current read-back signals are monitored through the digital system. Detailed information regarding these electronics can be found in [62].

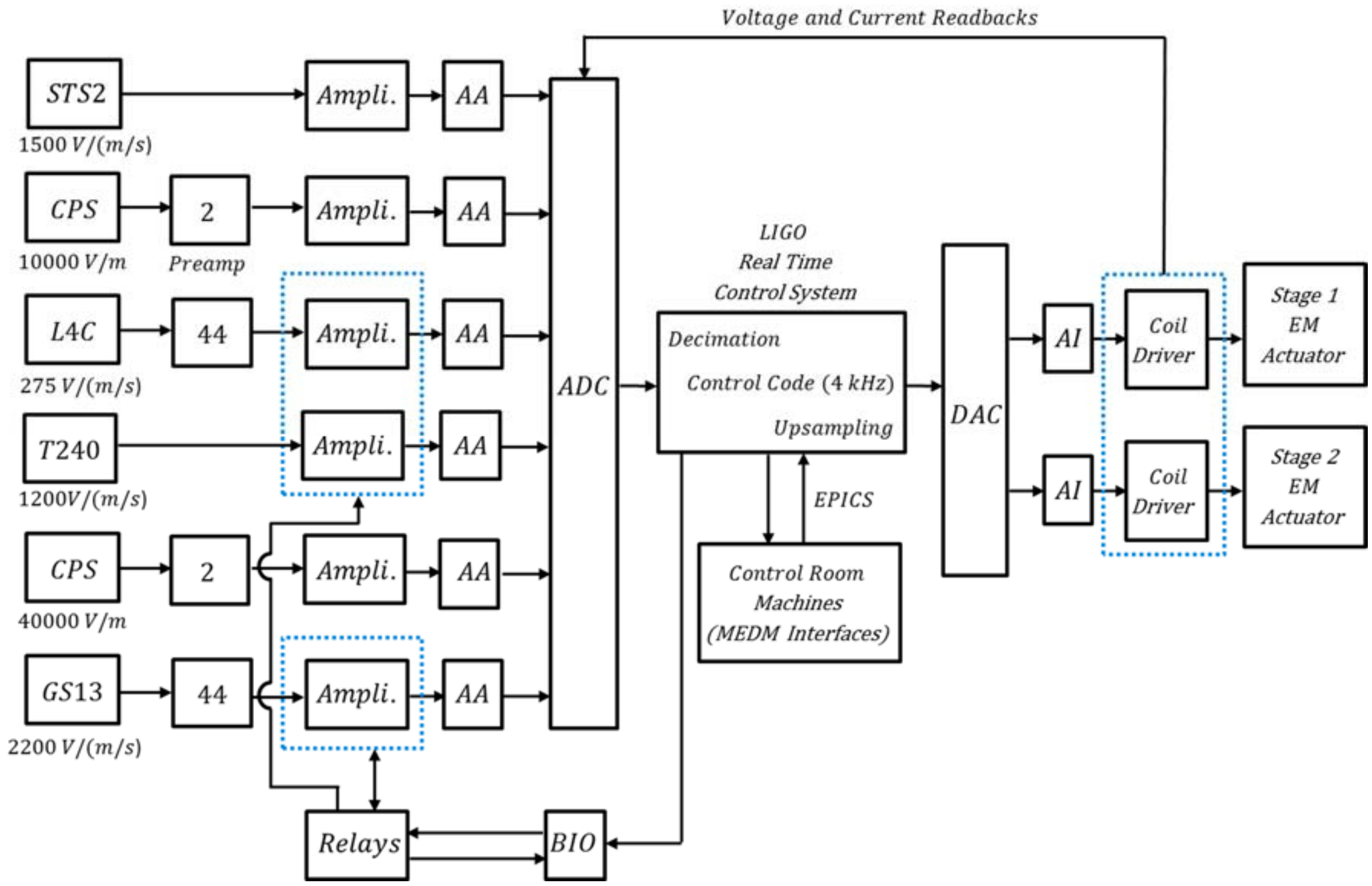

**Fig. 22 Simplified signal flow diagram.**

# 4 Design for series production

## *4.1 Production scale and Assembly considerations*

Unlike similar prototypes previously built, the BSC-ISI system has been engineered to allow timely production of a series of 15 units. As a result, a BSC-ISI unit can be fully assembled and tested in less than 4 weeks. It can be commissioned to operate near design sensitivity in about a week. Thirteen of the fifteen units necessary for the Advanced LIGO project have been built and tested (as of July 2014).

In order to reach the level of isolation described in the introduction, the system must operate in vacuum. All of the system's components (structure, instruments & cables) are ultra-high-vacuum compatible. They are subjected to a strict cleaning process that includes chemical cleaning, ultra-sound baths and low temperature heating in vacuum bake ovens. LIGO clean and bake procedures are detailed in [63]. The components are subjected to a thorough quality control process, including RGA scans and FTIR tests. The systems are assembled in Class 100 standard clean room. A BSC-ISI unit during the assembly process is shown in Fig. 23. The assembly process of the main structure and all the sub-assemblies is fully detailed in [64].

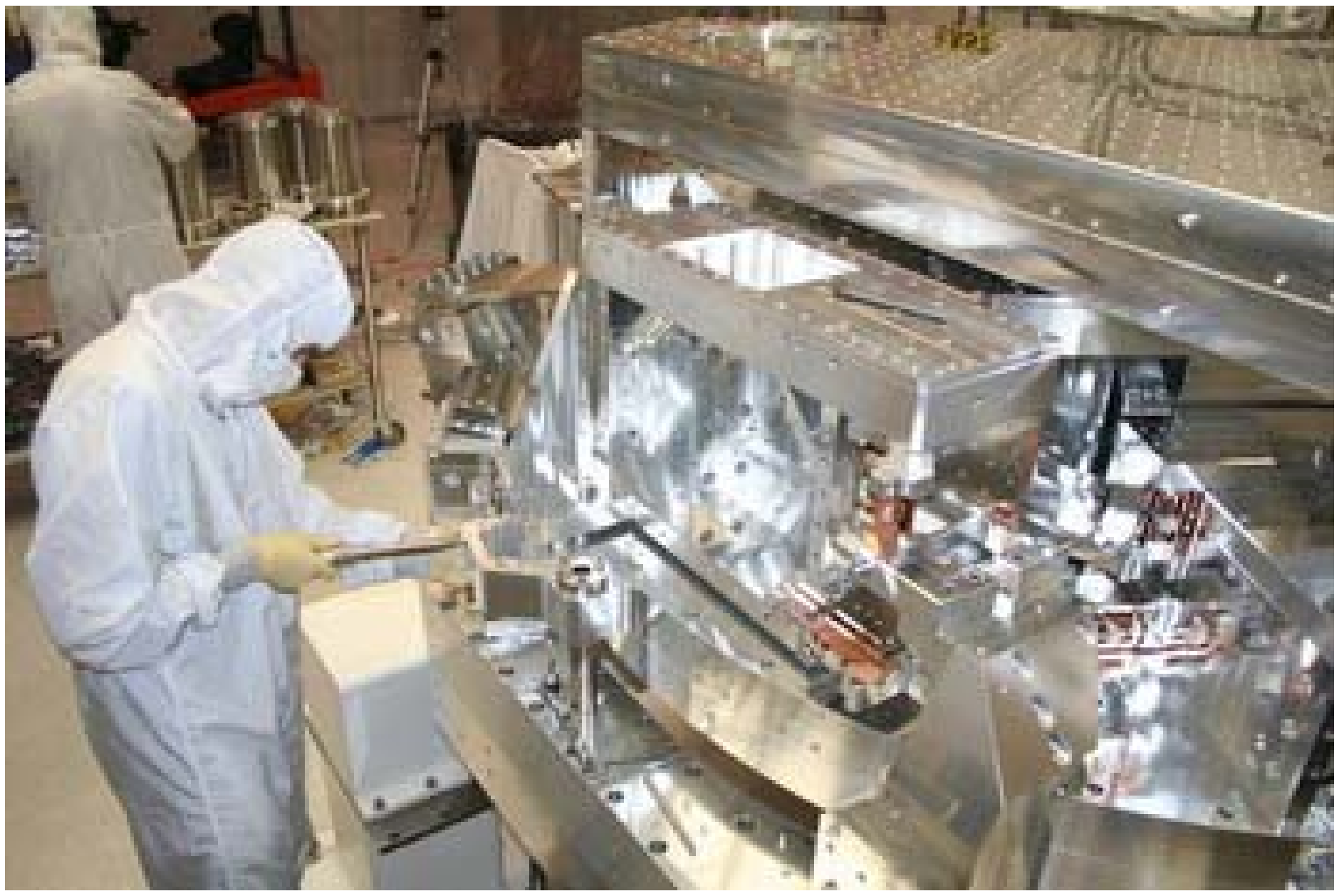

**Fig. 23 BSC-ISI Assembly in LIGO clean rooms.**

### *4.2 Testing process overview*

To reach the level of isolation targeted with this system, all the mechanical instruments and electronics components must operate at design sensitivity. The testing process is therefore a critical step of the system's integration phase. Before assembly, each component and sub-component such as actuators, sensors, pod assemblies, and electronic chassis is tested individually. After assembly, each BSC-ISI unit is submitted to a thorough testing process. The unit's testing procedure is detailed in [65]. It is made of three phases that can be summarized as follows.

The first phase of testing is dedicated to validate the BSC-ISI assembly. After assembly, the unit is loaded with dummy masses mounted on the top plate of stage 2 as shown in Fig. 5. The motion limiters are dis-engaged to release the two suspended stages. Due to machining and assembly tolerances, a little lateral motion is observed when the stages are released for the first time (typically a quarter of mm). The motion limiters are laterally re-aligned with this natural static equilibrium position. After that, the stages are leveled using trim masses attached to the sidewalls. The mass budget is carefully checked. The series of tests performed on each unit include geometry and leveling measurements, sensor noise tests, actuator response tests, range of motion measurements, system static response measurements, cross coupling measurements, linearity tests, transfer functions measurements, and damping loops tests.

After assembly testing, the dummy payload is removed. The BSC-ISI unit is transported from the assembly area to the detector area in a sealed container. The BSC-ISI is positioned on a stand designed for integration with the optical equipment. The equipment is installed on the optical table of the BSC-ISI. Fig. 24 shows a BSC-ISI supporting two multiple-pendulum suspensions. Great care must be given to the routing of the payload cables so as to not shortcut the BSC-ISI passive isolation. Fig. 25 shows how cables are routed from Stage 2 to Stage 1, and from stage 1 to Stage 0, to avoid direct coupling from Stage 0 to Stage 2 through the cabling. The cable parameters $k$ and $\eta$ symbolically represent the cable complex stiffness. Once the payload installation and cable routing are completed, the BSC-ISI is subjected to a second phase of testing called integration testing, during which the couplings between the platform and its payload are carefully checked.

After completion of the Integration testing, the entire cartridge (isolator and payload) is lifted and installed in the vacuum chamber. The third and last phase of testing is dedicated to the commissioning of the isolator in the vacuum environment. The digital controllers are installed and tuned to optimize the isolation performance of each unit.

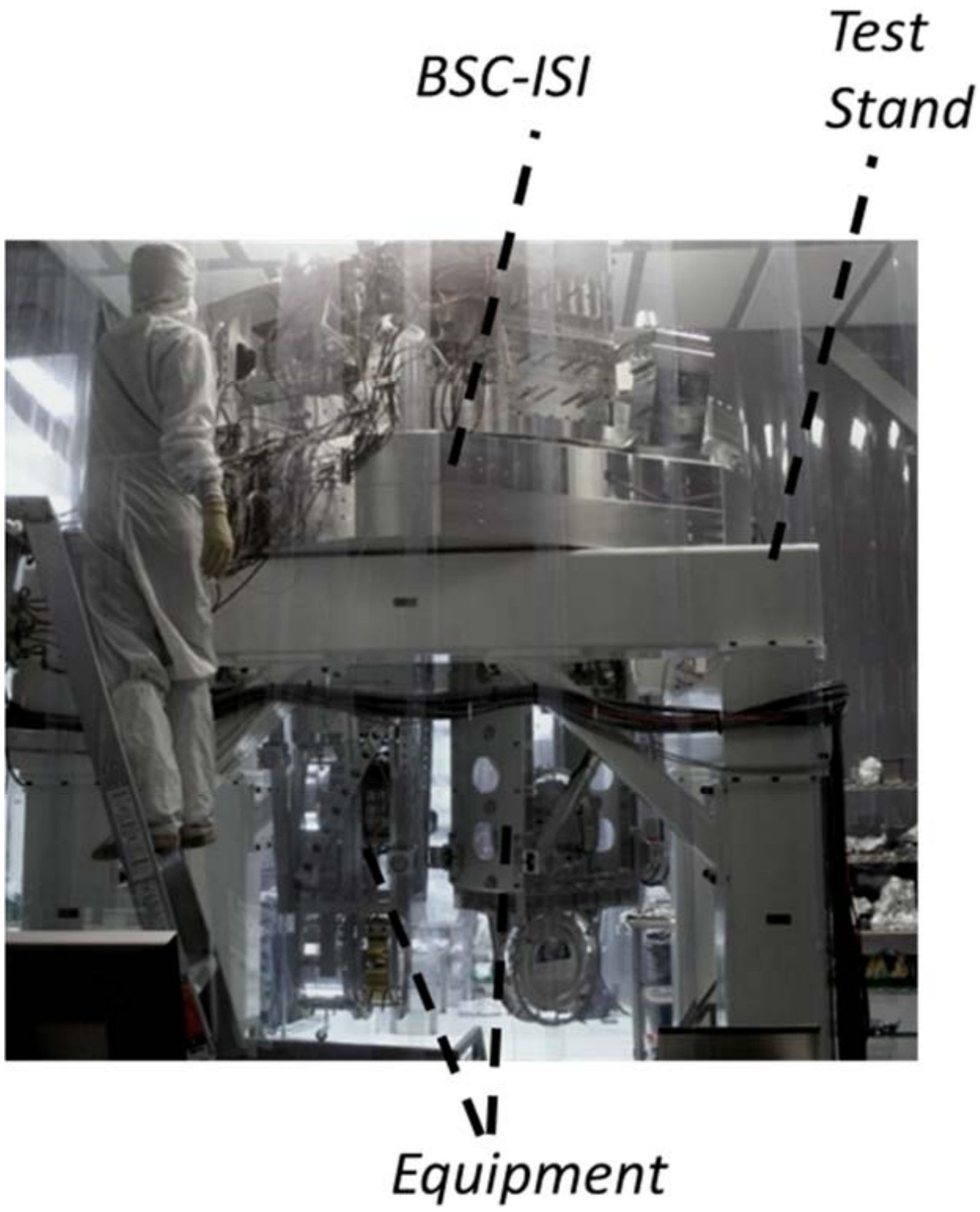

**Fig. 24 A BSC-ISI on integration stand.**

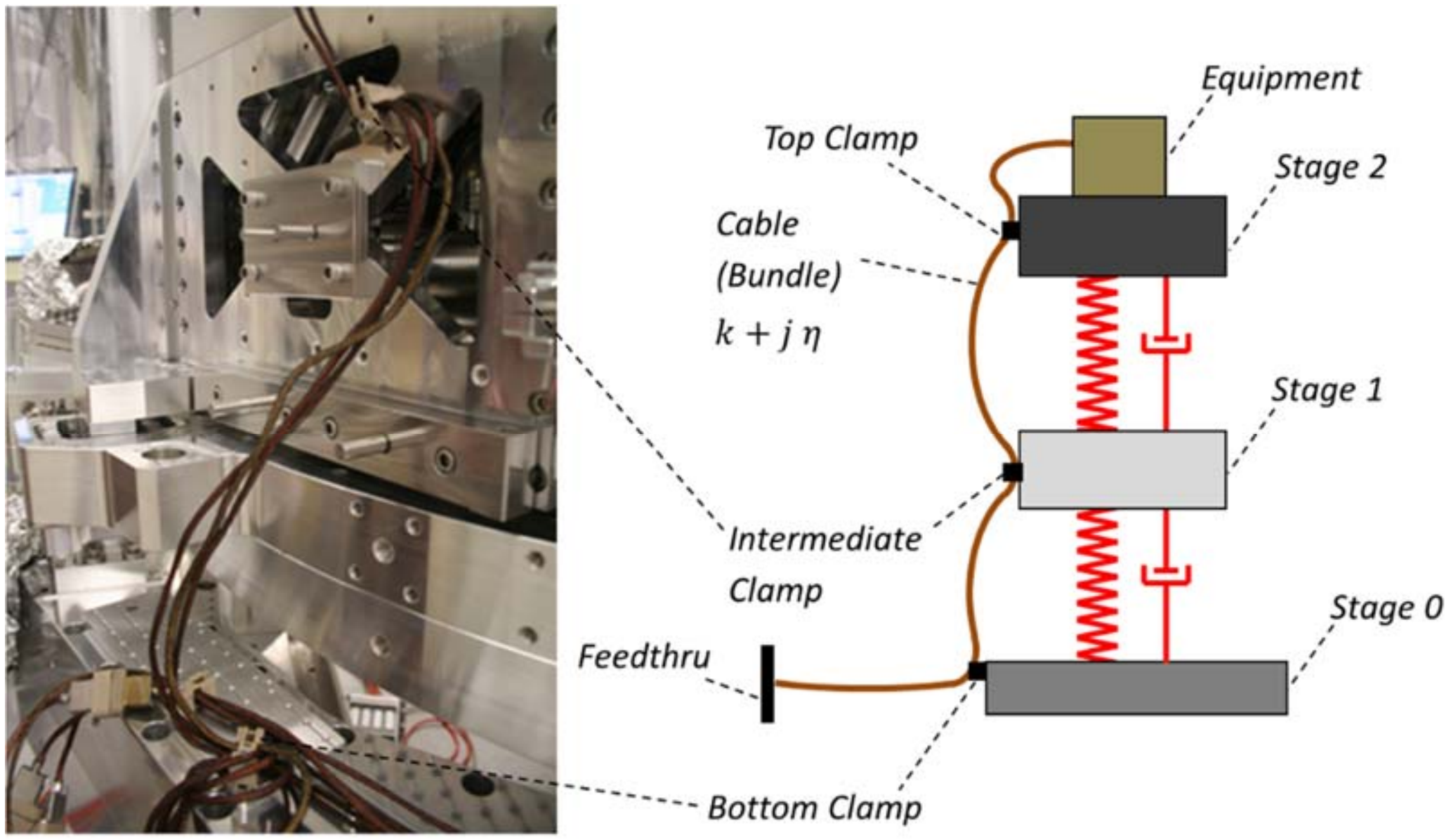

**Fig. 25 Cable routing.**

### *4.3 Production units frequency response*

In Fig. 26, the experimental transfer functions are compared with the response of an ideal two-mass spring system. Fig. 26 (a) shows the transfer function in the longitudinal ($X$) direction for stage 1. The actuators are combined to drive in the $X$ direction, and the L4Cs are used to sense the motion in the same direction. Below 1 Hz, the transfer function is very "flat". It indicates that there is very little tilt-horizontal coupling in static regime, and that the actuators are well aligned with respect to the system's static center of rotation. Between 1 Hz and 10 Hz, the main resonances are in agreement with the model. A little bit of coupling between $X$ and $RY$ is visible. It indicates that the center of mass is not perfectly aligned with the center of percussion, but it is sufficiently small so as to not affect the system's controllability and performance. The modal damping is induced by the magnetic actuators dissipating energy in the input resistance of the amplification stages. Between 10 Hz and 100 Hz, the system provides passive isolation as predicted. Above 150 Hz, the peaks of the structural resonances are damped to low level by passive dampers [48]. The Stage 2 horizontal transfer function show a similar set of characteristics, as shown in Fig. 26 (b). A notable difference is below 0.1 Hz, where the transfer function's corner frequency indicates higher tilt horizontal coupling ratio. This is caused by the rotational reaction of Stage 1. This coupling is greatly reduced when the rotational control of Stage 1 is engaged. Fig. 26 (c) and (d) show the transfer functions for the vertical directions. The good agreement with the model shows that there is very little cross-coupling between the degrees of freedom. The transfer functions measured in all other degrees of freedom are also in excellent agreement with the ideal two-degrees of freedom model response. The transfer function results are repeatable from one unit to another. Table 3 gives a summary of the system's rigid-body mode frequencies.

**Table 3: BSC-ISI Rigid-body modes frequencies.**

| Direction | In Phase Mode | Out of Phase Mode |
|---|---|---|
| X (or Y) | 1.25 Hz | 5.25 Hz |
| Z | 1.80 Hz | 6.70 Hz |
| RX (or RY) | 1.55 Hz | 6.80 Hz |
| RZ | 1.40 Hz | 5.70 Hz |

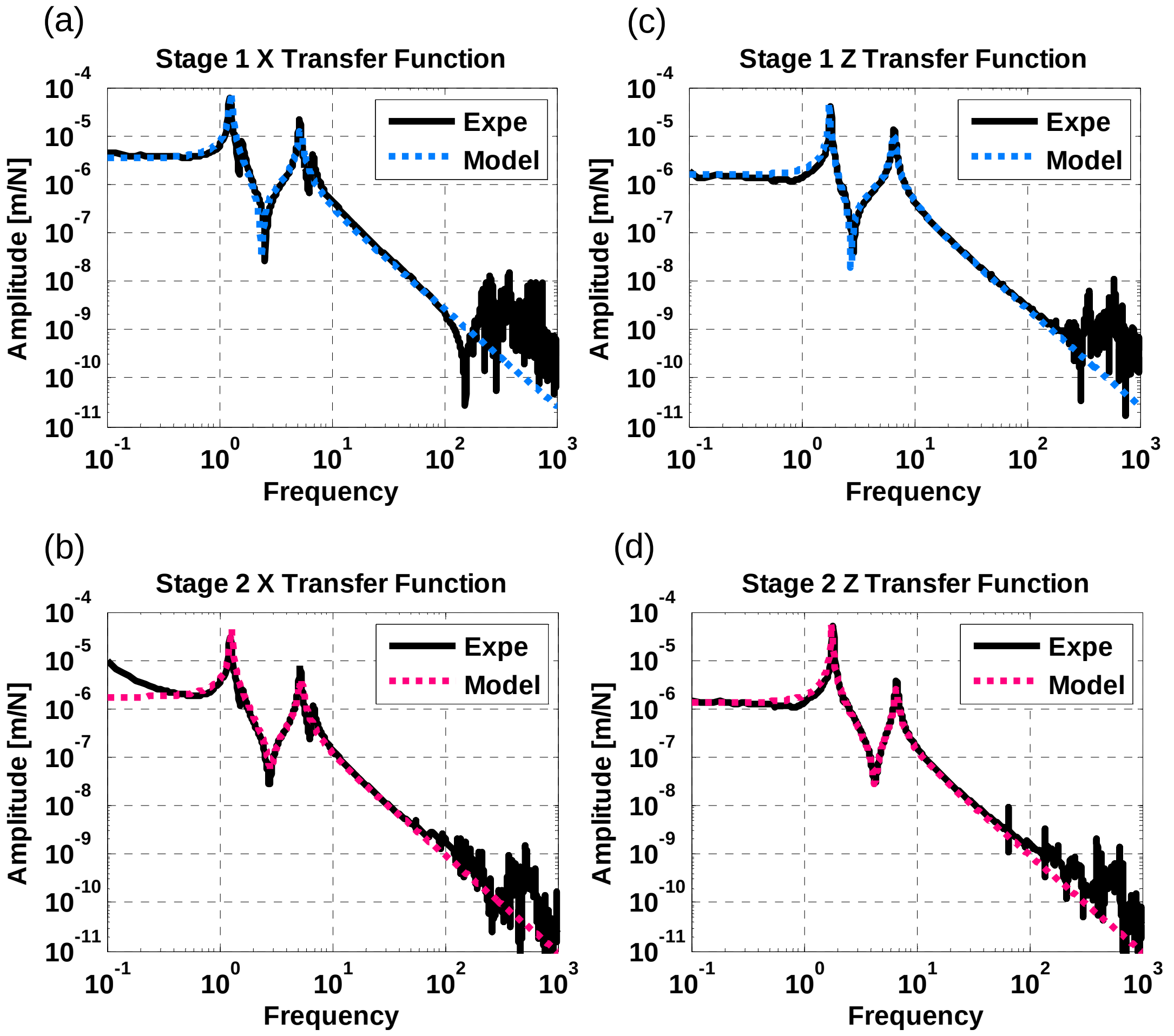

**Fig. 26 Comparison of theoretical transfer functions (Model) and experimental transfer functions as measured by the inertial sensors (Expe).**

## 5 Conclusion

An extensive engineering effort has been led during the past several years to develop the final version of the two-stage vibration isolation system for the Advanced LIGO observatories. The goal was to engineer a system not only to meet very high performance criteria, but also suitable for timely production, assembly, testing and commissioning for a series of 15 units.

A detailed presentation of the system's design has been made. It highlighted features of interest for the field of vibration isolation in precision engineering. It described the system architecture, the design of the flexure components, the actuators, the low-noise instrumentation, electronics, and challenges related to vacuum compatibility and the control strategy. It detailed how the system has been engineered for the production of a series of units.

During the past two years, 13 BSC-ISI units have been assembled and tested. The last two units are under construction and will be completed by the end of 2014. All the units tested show extremely reproducible results and characteristics. Five units are currently in use at each of the LIGO observatories and performing at very high level of isolation required for Advanced LIGO. A companion paper (Part 2: Experimental Investigation and Tests Results) covers in detail the results of the prototyping and testing phase of this project.

**Acknowledgments**

The authors acknowledge and gratefully thank the National Society Foundation for their support. LIGO was constructed by the California Institute of Technology and Massachusetts Institute of Technology with funding from the National Science Foundation and operates under cooperative agreement PHY-0107417.

We thank the JILA group for pioneering the work on active isolation systems using low frequency inertial sensors, and for demonstrating the feasibility of such multi-stage systems. We thank our colleagues from the suspension groups in GEO and LIGO for introducing us to the benefits of using triangular maraging steel blades to provide vertical isolation. We thank High Precision Devices for the mechanical design's realization of the rapid prototype and the technical demonstrator. We thank Alliance Space Systems Incorporation for the mechanical design's realization of the two-stage prototype. We thank Nanometrics, Streckeisen, Geotech, Sercel and Microsense for supplying us with great instruments, and for their technical support.

Finally yet importantly, this work would not have been possible without the outstanding support of the LIGO laboratory management, computer and data systems, procurement, facility modification and preparation, assembly and installation teams.

This document has been assigned LIGO Laboratory document number LIGO-P1200010.

## References

[1] Okano M, Kajimura K, Wakiyama S, Sakai F, Mizutani W, & Ono M. Vibration isolation for scanning tunneling microscopy. Journal of Vacuum Science & Technology A: Vacuum, Surfaces, and Films. 1987; 5.6; 3313-3320.

[2] Kuk Y, and Silverman PJ. Scanning tunneling microscope instrumentation. Review of scientific instruments. 1989; 60.2; 165-180.

[3] Rinker R. Super Spring - A New Type of Low-Frequency, Vibration Isolator. Ph.D. thesis. University of Colorado, Boulder. 1983.

[4] Nelson PG. An active vibration isolation system for inertial reference and precision measurement. Review of scientific instruments. 1991; 62.9; 2069-2075.

[5] Hensley JM, Peters A, & Chu S. Active low frequency vertical vibration isolation. Review of scientific instruments. 1999; 70.6; 2735-2741.

[6] Suzuki H, and Ho CM. A magnetic force driven chaotic micro-mixer. Micro Electro Mechanical Systems, 2002. The Fifteenth IEEE International Conference on. IEEE. 2002.

[7] Artoos K, Collette C, Brunetti L, Coe P, Geffroy N, Guinchard M, et al. Study of the stabilization to the nanometer level of Mechanical Vibrations of the CLIC Main Beam Quadrupoles. No. CERN-ATS-2009-127. 2009.

[8] Preumont A, Horodinca M, Romanescu I, De Marneffe B, Avraam M, Deraemaeker A, et al. A six-axis single-stage active vibration isolator based on Stewart platform. Journal of sound and vibration. 2007; 300.3; 644-661.

[9] Saulson PR. Vibration isolation for broadband gravitational wave antennas. Review of scientific instruments. 1984; 55.8; 1315-1320.

[10] Abramovici A, Althouse WE, Drever RW, Gürsel Y, Kawamura S, Raab FJ, et al. LIGO: The laser interferometer gravitational-wave observatory. Science. 1992; 256.5055; 325-333.

[11] Robertson NA, Drever RWP, Kerr I, & Hough J. Passive and active seismic isolation for gravitational radiation detectors and other instruments. Journal of Physics E: Scientific Instruments. 1982; 15.10; 1101.

[12] Abbott BP, Abbott R, Adhikari R, Ajith P, Allen B, Allen G, et al. LIGO: the laser interferometer gravitational-wave observatory. Reports on Progress in Physics. 2009; 72.7; 076901.

[13] Willke B, Aufmuth P, Aulbert C, Babak S, Balasubramanian R, Barr BW, et al. The GEO 600 gravitational wave detector. Classical and Quantum Gravity. 2002; 19.7; 1377.

[14] Bradaschia C, Del Fabbro R, Di Virgilio A, Giazotto A, Kautzky H, Montelatici V, et al. The VIRGO project: a wide band antenna for gravitational wave detection. Nuclear Instruments and Methods in Physics Research Section A: Accelerators, Spectrometers, Detectors and Associated Equipment. 1990; 289.3; 518-525.

[15] Ando M, & TAMA collaboration. Current status of the TAMA300 gravitational-wave detector. Classical and Quantum Gravity. 2005; 22.18; S881.

[16] Kuroda K. Status of LCGT. Classical and Quantum Gravity. 2010; 27.8; 084004.

[17] Plissi MV, Torrie CI, Husman ME, Robertson NA, Strain KA, Ward H, et al. GEO 600 triple pendulum suspension system: Seismic isolation and control. Review of scientific instruments. 2000; 71.6; 2539-2545.

[18] Grote H, and LIGO Scientific Collaboration. The status of GEO 600. Classical and Quantum Gravity. 2008; 25.11; 114043.

[19] Aston SM, Barton MA, Bell AS, Beveridge N, Bland B, Brummitt AJ, et al. Update on quadruple suspension design for Advanced LIGO. Classical and Quantum Gravity. 2012; 29.23; 235004.

[20] Losurdo G, Calamai G, Cuoco E, Fabbroni L, Guidi G, Mazzoni M, et al. Inertial control of the mirror suspensions of the VIRGO interferometer for gravitational wave detection. Review of Scientific Instruments. 2001; 72.9; 3653-3661.

[21] Accadia T, Acernese F, Antonucci F, Astone P, Ballardin G, Barone F, et al. Status of the Virgo project. Classical and Quantum Gravity 2011; 28.11; 114002.

[22] Acernese F, Antonucci F, Aoudia S, Arun KG, Astone P, Ballardin G, et al. Measurements of Superattenuator seismic isolation by Virgo interferometer. Astroparticle Physics. 2010; 33.3; 182-189.

[23] Somiya K. Detector configuration of KAGRA–the Japanese cryogenic gravitational-wave detector. Classical and Quantum Gravity. 2012; 29.12; 124007.

[24] Harry GM. Advanced LIGO: the next generation of gravitational wave detectors. Classical and Quantum Gravity. 2010; 27.8; 084006.

[25] Abbott R, Adhikari R, Allen G, Cowley S, Daw E, DeBra D et al. Seismic isolation for Advanced LIGO. Classical and Quantum Gravity. 2002; 19.7; 1591.

[26] Abbott R, Adhikari R, Allen G, Baglino D, Campbell C, Coyne D, et al. Seismic Isolation Enhancements for Initial and Advanced LIGO. Class. Quantum Grav. 2004; 21; 915-921.

[27] Robertson NA, Abbott B, Abbott R, Adhikari R, Allen GS, Armandula H, et al. Seismic Isolation and Suspension Systems for Advanced LIGO. Gravitational Wave and Particle Astrophysics Detectors. Proceedings of SPIE. 2004.

[28] Hardham CT, Allen GS, DeBra DB, Hua W, Lantz BT, Nichol JG. Quiet Hydraulic Actuators for the Laser Interferometer Gravitational Wave Observatory (LIGO). Proceedings of ASPE conference on Control of Precision Systems. 2001.

[29] Hardham CT, Abbott B, Abbott R, Allen G, Bork R, Campbell C, et al. Multi-DOF Isolation and Alignment with Quiet Hydraulic Actuators Control of Precision Systems. Proceedings of ASPE conference on Control of Precision Systems. 2004.

[30] Hua W, Adhikari R, DeBra DB, Giaime JA, Hammond GD, Hardham CT, et al. Low frequency active vibration isolation for Advanced LIGO. Proc. of SPIE Vol. Vol. 5500. 2004.

[31] Wen S, Mittleman R, Mason K, Giaime JA, Abbott R, Kern J, O'Reilly B, et al. Hydraulic External Pre-Isolator System for LIGO. arXiv preprint arXiv:1309.5685 (2013).

[32] Matichard F, et al. LIGO Vibration Isolation and Alignment Platforms: an Overview of Systems, Features and Performance of Interest for the Field of Precision Positioning and Manufacturing. In Proceedings of ASPE conference on Precision Control for Advanced Manufacturing Systems. 2013.

[33] Danaher C, Hollander B, et al. Advanced LIGO Single Stage HAM Vibration Isolation Table. LIGO document LIGO-G070156 (2007).

[34] Kissel JS. Calibrating and Improving the Sensitivity of the LIGO Detectors. Diss. Louisiana State University. 2010.

[35] Matichard F, Abbott B, Abbott S, Allewine E, Barnum S, Biscans S. et al. Prototyping, Testing, and Performance of the Two-Stage Seismic Isolation System for Advanced LIGO Gravitational Wave Detectors. In Proceedings of ASPE conference on Control of Precision Systems. 2010.

[36] Fritschel P. Seismic isolation subsystem design requirement. LIGO document E990303. 2001.

[37] Stebbins RT, Newell D, Richman SN, Bender PL, Faller JE, et al. Low-frequency active vibration isolation system. Proc. SPIE. Vol. 2264. 1994.

[38] Richman SJ, Giaime JA, Newell DB, Stebbins RT, Bender PL, et al. Multistage active vibration isolation system. Review of Scientific Instruments. 1998; 69.6; 2531-2538.

[39] Newell DB, Richman SJ, Nelson PG, Stebbins RT, Bender PL, Faller JE, & Mason J. An ultra-low-noise, low-frequency, six degrees of freedom active vibration isolator. Review of scientific instruments. 1997; 68.8; 3211-3219.

[40] Giaime JA, et al. Baseline LIGO-II implementation design description of the stiff active seismic isolation system. LIGO document T000024. 2000.

[41] Giaime JA, et al. Advanced LIGO Seismic Isolation System Conceptual Design. LIGO document E010016. 2001.

[42] Lantz B, Lessons from the ETF Technology Demonstrator, LIGO document G050271. 2005.

[43] Coyne D, & al, Design Requirements for the In-Vacuum Mechanical Elements of the Advanced LIGO Seismic Isolation System for the BSC Chamber. LIGO Document E030179. 2004.

[44] Smith K, Advanced LIGO BSC Prototype Critical Design Review. ASI Document 20008644. 2004.

[45] Smith K, Post-CDR Design Assessments of BSC Structure. ASI Technical Memorandum 20009033. 2004.

[46] Matichard F, et al. Advanced LIGO Preliminary Design Review of the BSC ISI system. LIGO Document L0900118. 2009.

[47] Matichard F, Mason K, Mittleman R, Lantz B, Abbott B, MacInnis M, et al. Dynamics Enhancements of Advanced LIGO Multi-Stage Active Vibration Isolators and Related Control Performance Improvement. In ASME 2012 International Design Engineering Technical Conferences and Computers and Information in Engineering Conference (pp. 1269-1278). American Society of Mechanical Engineers. 2012.

[48] Matichard, F., et al. LIGO Two-Stage Vibration Isolation and Positioning Platform, Part 2: Experimental Investigation and Tests Results. Submitted for publication to Precision Engineering. 2014.

[49] Matichard F, et al. aLIGO BSC-ISI, Top and Sub-Assemblies Drawings. LIGO document E1000397. 2010.

[50] Matichard F, et al. ISI actuator dynamic response identification. LIGO document T0900226. 2009.

[51] Hilliard M, et al. aLIGO BSC-ISI, General Assembly Procedure. LIGO document E0900357. 2009.

[52] Clark D. Replacement Flexures for the GS-13 Seismometer. LIGO document T0900089. 2009.

[53] Lantz B. LT1012 is the best op-amp for the GS13 preamp. LIGO document T0900457. 2009.

[54] Stein A, FEA of BSC ISI Springs. LIGO document T0900569. 2009.

[55] Clark D. Damping Structures - Poster for LVC Arcadia Meeting. LIGO document G1000237. 2010.

[56] Smith K, Analysis of LIGO Flexures Rods. ASI Technical Note 20007235. 2004.

[57] Rodgers, Peter W. Maximizing the signal-to-noise ratio of the electromagnetic seismometer: The optimum coil resistance, amplifier characteristics, and circuit. Bulletin of the Seismological Society of America 83.2 (1993): 561-582.

[58] Rodgers, P. W. (1994). Self-noise spectra for 34 common electromagnetic seismometer/preamplifier pairs. Bulletin of the Seismological society of America, 84(1), 222-228.

[59] Bork R, et al. New control and data acquisition system in the Advanced LIGO project. Proc. of Industrial Control And Large Experimental Physics Control System (ICALEPSC) conference. 2011.

[60] Dalesio LR, et al. The experimental physics and industrial control system architecture: past, present, and future. Nuclear Instruments and Methods in Physics Research Section A: Accelerators, Spectrometers, Detectors and Associated Equipment. 1994; 352.1; 179-184.

[61] Evans K. MEDM Reference Manual. Argonne National Laboratory. 2005.

[62] Abbott B. SEI Electronics Document Hub, LIGO document T1300173. 2013.

[63] Bland B, Coyne D, Fauver J. LIGO Clean and Bake Methods and Procedures. LIGO document E960022-v25. 2013.

[64] Hillard M, Le Roux A, & al. aLIGO BSC-ISI, General Assembly Procedure. LIGO Document E0900357-v28. 2012.

[65] Matichard F, Lantz B, et al. aLIGO BSC-ISI Testing and Commissioning Documentation. LIGO Document E1000306. 2013.

## List of Figures

Fig. 14. Stage 1-2 actuator assembly.

Fig. 15. Stage 1-2 vertical actuator CAD representation.

Fig. 16. GS13 equipped with mu-metal shielding and sealed in a custom pod for use in ultra-high-vacuum.

Fig. 17. Conceptual representation of the spring assembly.

Fig. 18. Stage 0-1 spring assembly.

Fig. 19. Stage 1-2 spring assembly.

Fig. 20. Clamp gap inspection: a gap between the blade and its mount can arise if the load force over comes the pre-load force in the bolted assembly. Friction in the ultra-clean bolted assembly and pre-load are carefully controlled.

Fig. 21. (a) Conceptual representation of the flexure rod attached to the blade. (b) Force and position parameters.

Fig. 22 Simplified signal flow diagram.

Fig. 23 BSC-ISI Assembly in LIGO clean rooms.

Fig. 24 A BSC-ISI on integration stand.

Fig. 25 Cable routing.

Fig. 26 Comparison of theoretical transfer functions (Model) and experimental transfer functions as measured by the inertial sensors (Expe).

## List of Tables